\documentclass{emulateapj}
\slugcomment{{\sc Accepted to ApJ:} May 11, 2006} 
\usepackage{amsmath}

\def\ni{\noindent}
\def\s{{\rm\,s}}
\def\erg{{\rm\,erg}}
\def\cm{{\rm\,cm}}
\def\m{{\rm\,m}}
\def\km{{\rm\,km}}

\def\gm{{\rm\,g}}
\def\g{{\rm\,g}}

\def\AU{{\rm\, AU}}
\def\mum{\,\mu{\rm m}}

\def\yr{{\rm\,yr}}

\def\pomega{\widetilde{\omega}}

\begin{document}

\shortauthors{}
\shorttitle{AU Mic Disk}


\title{Dust Dynamics, Surface Brightness Profiles, and Thermal Spectra of
Debris Disks:
 The Case of AU Mic}

\author{Linda E. Strubbe\altaffilmark{1} 
\& Eugene I. Chiang\altaffilmark{1,2}} 
\altaffiltext{1}{Center for Integrative Planetary Sciences,
Astronomy Department,
University of California at Berkeley,
Berkeley, CA~94720, USA}

\altaffiltext{2}{Alfred P.~Sloan Research Fellow}

\email{linda@astron.berkeley.edu}

\begin{abstract}
AU Microscopii is a 12 Myr old M dwarf that harbors an optically thin,
edge-on disk of dust.  The scattered light surface brightness falls
with projected distance $b$ from the star as $b^{-\alpha}$;
within $b = 43\,\,\rm{AU}$, $\alpha \approx 1$--2, while outside 43 AU,
$\alpha \approx 4$--5. We devise a theory to explain this profile. At
a stellocentric distance $r = r_{\rm BR} = 43\,\,\rm{AU}$, we posit a
ring of parent bodies on circular orbits: the ``birth ring,'' wherein
micron-sized grains are born from the collisional attrition of parent
bodies. The ``inner disk'' at $r < r_{\rm BR}$ contains grains that
migrate inward by corpuscular and Poynting-Robertson (CPR) drag. The
``outer disk'' at $r > r_{\rm BR}$ comprises grains just large enough
to remain bound to the star, on orbits rendered highly eccentric by
stellar wind and radiation pressure.  How the vertical optical depth
$\tau_\perp$ scales with $r$ depends on the fraction of grains that
migrate inward by CPR drag without suffering a collision.  If this fraction is
large, the inner disk and birth ring share the same optical depth, and
$\tau_\perp \propto r^{-5/2}$ in the outer disk. By contrast, under
collision-dominated conditions, the inner disk is empty, and
$\tau_\perp \propto r^{-3/2}$ outside. These scaling relations, which
we derive analytically and confirm numerically, are robust against
uncertainties in the grain size distribution. By simultaneously
modeling the surface brightness and thermal spectrum, we break model
degeneracies to establish that the AU Mic system is
collision-dominated, and that its narrow birth ring contains a lunar
mass of decimeter-sized bodies.  The inner disk is devoid of
micron-sized grains; the surface brightness at $b \lesssim
43\,\,\rm{AU}$ arises from light forward scattered by the birth
ring. Inside $b = 43\,\,\rm{AU}$, the disk's $V-H$ color should not
vary with $b$; outside, the disk must become bluer as ever smaller
grains are probed.

\end{abstract}


\section{INTRODUCTION}
\label{intro}

``Debris disks'' surrounding young stars are 
composed of optically thin dust (see the
reviews by Artymowicz 2000; 
Lagrange, Backman, \& Artymowicz 2000; Zuckerman 2001).
Most debris disks are inferred to exist from measurements of infrared
excesses (e.g., Aumann et al.~1984; Habing et al.~2001; 
Zuckerman \& Song 2004).
A few disks are close enough to resolve in images, 
either in scattered starlight
(e.g., Smith \& Terrile 1984; Schneider et al.~1999; 
Kalas, Graham, \& Clampin~2005)
or in thermal emission (e.g., Telesco et al.~2000; Greaves et
al.~2004).

Is the observed dust primordial---the remains of an optically thick,
gas-rich disk from a previous Herbig Ae or T Tauri phase?
Or is it maintained in equilibrium---continuously
removed by processes such as Poynting-Robertson drag
and replenished by the comminution of larger, colliding parent bodies?
A third possibility is that the observed dust 
represents the transient aftermath of recent
cataclysmic events. Dust might be freshly generated, 
unequilibrated debris from the catastrophic
destruction of large planetesimals (Su et al.~2005; Song et al.~2005).


The debris disk encircling the young M dwarf AU Microscopii 
is a promising place to
investigate these questions. It is well resolved
in scattered light from optical to near-infrared 
wavelengths (Kalas, Liu, \& Matthews 2004; Krist et al.~2005, hereafter K05;
Liu 2004; Metchev et al.~2005).
Of central relevance to our study is the disk's surface brightness profile.
Within a projected distance $b$ from the star of 43 AU, the surface brightness
SB falls approximately as $b^{-1.8}$ 
(K05). We refer to this region
as the ``inner disk.''  Outside 43 AU, in the ``outer disk,'' the slope
of the profile changes dramatically: SB $\propto b^{-4.7}$ (K05).
This break is observed independently by other researchers 
(Liu 2004; Metchev et al.~2005).
The shape of the profile is all the more significant because it resembles
that of the debris disk surrounding $\beta$ Pictoris
\citep{kalasjewitt,liu}, and of the recently discovered disk
encompassing HD 139664
\citep{kalas2006}.  AU Mic's disk is also detected in unresolved thermal
emission \citep{liu2, chen}.  The disk's infrared spectrum peaks at a
wavelength of
$\sim$$100\mum$ and exhibits no excess at $12\mum$; this behavior
suggests that the disk contains an
inner hole \citep{liu2}.

Here we offer a theory that explains these observations quantitatively.
The reason why the surface brightness profile breaks at 43 AU is 
that a narrow ring of parent bodies,
analogous to the solar system's Kuiper 
Belt, orbits the star
at a stellocentric radius $r = r_{\rm BR} = 43 \AU$.  
The subscript ``BR'' refers to our term for the
belt of parent bodies, the ``birth ring,''
wherein micron-sized dust grains are born through collisions of
larger planetesimals.
Grain creation is balanced in steady state by
destructive
grain-grain collisions and removal by corpuscular and Poynting-Robertson (CPR) 
drag.\footnote{Our analysis ignores
any gas that might still be orbiting the star. Only upper limits are
observed for
the column of gas towards the star: $N_{\rm H_2} <
1.7\times10^{19}\cm^{-2}$ \citep{roberge}.}
Corpuscular drag exerted by the young M dwarf's wind
is probably at least a few times
more effective at removing
dust than Poynting-Robertson drag in the AU Mic system, a possibility
first pointed out by
Plavchan, Jura, \& Lipscy (2005, hereafter PJL05).
The outer disk comprises grains that are only tenuously bound,
moving on orbits rendered highly eccentric by stellar wind
and radiation forces (e.g., Lecavelier des Etangs et al.~1996;
Augereau et al.~2001). These barely bound grains dominate
scattering of starlight in the outer disk. 
By contrast, unbound grains escape the system too
quickly for their steady-state population to contribute appreciably
to the surface brightness.
The inner disk is populated by
grains that migrate inward by
CPR drag quickly enough to evade collisional destruction.
In CPR-dominated (what we refer to as ``type A'') disks,
a large fraction of grains meets
this condition,
unlike in collision-dominated (``type B'') disks.
Similar classifications were put forward by \citet{wyatt} and
\citet{meyer} in their considerations of disks composed of  
single-sized grains.
We calculate simultaneously
the steady-state spatial and size distributions of dust particles, and
derive how the outer disk's optical depth scales with radius
for type A and type B disks.  
Our analysis accounts for destructive grain-grain collisions and
the detailed dynamics of CPR drag, which reduces
not only the orbital semi-major axes of grains but also their orbital
eccentricities
(Wyatt \& Whipple 1950). The reduction of orbital eccentricity is not often
considered but is a key component of our theory.  
Whether type A or type B conditions apply to AU Mic's disk
is determined in part by the strength of the
stellar wind, which according to previous works is 
uncertain by orders of magnitude.
In this work, we place a novel constraint on the stellar mass loss
rate and decide the appropriate disk type by comparing our theoretical
models to the observations.

In \S\ref{oom}, we lay down basic parameters of the AU Mic system:
stellar properties, disk optical depths, timescales for
grain-grain collisions, and how the star's wind and radiation
alter orbits of dust grains. In \S\ref{theory}, 
we employ order-of-magnitude physics
and analytic scalings to understand how the interplay between collisions,
blow-out, and drag shapes the observed surface brightness 
profiles of the inner and outer disks.
There we derive the steady-state grain size distribution as a function
of position, including the maximum sizes and total mass of parent bodies.
In \S\ref{montecarlo}, we verify and extend our analytic results with
a Monte Carlo simulation of the disk's surface brightness, color, and
spectral energy distribution (SED).
Models are compared directly with observations.
Finally, in \S\ref{discussion}, we summarize our theory,
place it in context with our understanding of how planets form,
and point out directions for future research.

\section{PRELIMINARIES}
\label{oom}

We establish orders of magnitude characterizing the AU Mic system.
Stellar properties---including the stellar mass-loss rate that
figures prominently throughout our analysis---are 
reviewed in \S\ref{starprop}.
Collision times
between grains and relative collision velocities are estimated in
\S\ref{coltimevis}. Grain dynamics relevant to our theory include
blow-out by stellar wind and radiation pressure, as discussed in
\S\ref{blowout},
and orbital decay by corpuscular and Poynting-Robertson drag,
as treated in \S\ref{cprdrag}.

\subsection{Stellar Properties}
\label{starprop}

AU Microscopii is a spectral type dM1e
star located a distance $d = 9.9$ pc from Earth.
It has mass $M_{\ast}=0.5
M_{\odot}$, radius $R_{\ast} =
0.93 R_{\odot}$, effective temperature $T_{\ast} = 3500$ K,
luminosity $L_{\ast}=0.1 L_{\odot}$, $V$ magnitude 8.8, 
and $H$ magnitude 4.8
(Kalas et al.~2004; Metchev et al.~2005; and
references therein).
The star's age is
estimated to be $t_{\rm age} = 12^{+8}_{-4}$ Myr.
AU Mic's X-ray luminosity is a prodigious $L_{\rm X} = 5.5 \times 10^{29} \,
{\rm erg/s} = 3\times 10^{-3} L_{\ast}$
\citep{hunsch}.  The star flares at both X-ray and ultraviolet
wavelengths \citep{magee}.  The stellar
rotation period is 4.87 days \citep{torres}.

%

How strong is AU Mic's stellar wind?
The wind velocity $v_{\rm wind}$ is likely of
order the stellar escape velocity $v_{\rm esc} =
\sqrt{2GM_{\ast}/R_{\ast}} \approx 450 {\rm\, km/s}$.
Plavchan et al.~(2005) discuss what is known about mass loss
rates $\dot{M}_{\ast}$ from M dwarfs, citing values ranging
from 10 to as high as 10$^4$ times the solar mass-loss rate of
$\dot{M}_{\odot} = 2 \times 10^{-14} M_{\odot} \yr^{-1}$.
While the star's youth, flaring activity, and fast
rotation suggest that a powerful wind emanates from AU Mic,
the star's X-ray emission indicates otherwise.
Wood et al.~(2002) and Wood et al.~(2005) 
study the relationship between X-ray luminosity and stellar mass loss
rate by measuring $\dot{M}_\ast$ from a handful of M, K, and G dwarfs having
ages $\gtrsim 500$ Myr.  They establish that the mass flux
at the stellar surface $F_M \equiv \dot{M}_{\ast}/(4\pi R_{\ast}^2)$
increases with X-ray surface flux
$F_{\rm X}\equiv L_{\rm X}/(4\pi R_{\ast}^2)$ for
$F_{\rm X} < 8 \times 10^5\erg\cm^{-2}\s^{-1} \equiv F_{\rm X, crit}$.
The mass flux $F_{\rm M}$ saturates at a maximum
of $10^2 F_{M,{\odot}}$. For $F_{\rm X} > F_{\rm X, crit}$, $F_M$ drops to
$\lesssim 10 F_{M,{\odot}}$,
perhaps because the strong magnetic
fields of such extraordinarily X-ray-active stars inhibit stellar winds
(see also Schrijver \& Title 2001; Strassmeier 2002).
AU Mic's
X-ray flux $F_{\rm X} \sim 1 \times 10^7 \erg\cm^{-2}\s^{-1}$ exceeds
$F_{\rm X, crit}$, implying at face value a relatively low mass-loss rate.
Nevertheless, it is unclear whether the measurements of Wood et al.~(2005)
apply to this highly variable, pre-main-sequence star.
To accommodate the uncertainty in AU Mic's mass-loss rate, 
we consider in our analysis a wide
range of values, $\dot{M}_{\ast} \in
(1,10,10^2,10^3) \dot{M}_{\odot}$.
We ultimately find in
\S\ref{montecarlo} that detailed comparison between theoretical models
of the disk and observations can, in fact, constrain
$\dot{M}_\ast$.


\subsection{Collision Times}
\label{coltimevis}

Consider for the moment
a disk of single-sized particles on low-eccentricity orbits. 
Where the vertical, geometric
optical depth equals $\tau_{\perp}$, the mean free time between collisions is

\begin{equation}
t_{\rm c} \sim \frac{1}{\Omega \tau_{\perp}} \,,
\label{tcol}
\end{equation}

\ni where $\Omega$ is the local orbital angular frequency. This expression
may be derived by recognizing that every $\sim$$1/\Omega$ orbital period,
a typical
particle traverses an optical depth of $\sim$$\tau_{\perp}$ over the course
of its vertical epicycle. For $\tau_{\perp} < 1$, the particle collides
with probability $\tau_{\perp}$; for $\tau_{\perp} > 1$, it undergoes
$\tau_{\perp}$ collisions per orbit.  

Detailed fits to observations in \S\ref{montecarlo} reveal
that the total vertical, geometric optical depth in the birth
ring at $r = r_{\rm BR} = 43\AU$ equals $\tau_{\perp,{\rm BR}} \equiv
\tau_{\perp} (r_{\rm BR}) = 4 \times 10^{-3}$. We define a fiducial
collision time

\begin{equation}
t_{{\rm c,BR}} \equiv \frac{1}{\Omega \tau_{\perp,{\rm BR}}} 
\sim 2 \times 10^4 \left(\frac{4\times10^{-3}}{\tau_{\perp,{\rm BR}}}
\right)
 \yr \,. 
\label{tcolvis}
\end{equation}


The collision lifetime $t_{\rm col}(s)$ is the time a grain of radius
$s$ survives before it is destroyed by colliding with another grain.
Unlike the case for
$t_{\rm c}$, in calculating
$t_{\rm col}$
we do not assume that particles have a single size.  For a given
collisional specific energy $Q^{\ast}$ 
\citep[ergs g$^{-1}$][]{greenberg,fujiwara}, 
targets of size $s$ suffer catastrophic
dispersal by smaller projectiles of minimum size

\begin{eqnarray}
s_{\rm proj} & \sim & \left( \frac{2Q^{\ast}}{v_{\rm rel}^2}
\right)^{1/3} s \nonumber \\ 
& \sim & 0.6 \, \left( \frac{Q^{\ast}}{10^7
  \erg \gm^{-1}} \right)^{1/3} \left( \frac{100 \m \s^{-1}}{v_{\rm
    rel}} \right)^{2/3} s \,,
\label{sproj}
\end{eqnarray}

\ni which is comparable to $s$.
By catastrophic dispersal we mean that the mass of 
the largest postcollision fragment
is no greater than half the mass of the original target and
that collision fragments disperse without gravitational reassembly.
We have normalized the relative speed $v_{\rm rel}$
to the vertical velocity dispersion of visible grains 
at $r = r_{\rm BR} = 43 \AU$,

\begin{equation}
\Omega \frac{h_{\rm BR}}{2} \sim 100 \m \s^{-1} \,,
\end{equation}

\ni where the full vertical disk height $h_{\rm BR} \approx 2.75\AU$ (K05).
We have normalized $Q^{\ast}$
to a value appropriate for centimeter-sized
silicate targets \citep{greenberg,fujiwara}.
Ice targets of similar size have specific energies
that are 2 orders of magnitude smaller \citep{greenberg}.
On the other hand, it is possible that $Q^{\ast}$ increases with decreasing
size \citep{fujiwara,housen},
perhaps as fast as $Q^{\ast} \propto s^{-0.5}$.
If so, grains having sizes $s \sim 1 \mum$ would be
$\sim$100 times stronger than their centimeter-sized counterparts,
thereby possibly cancelling the reduction in strength due to
an icy composition.
To keep the telling of our story as simple as possible,
we adhere to a nominal,
size-independent value of
$Q^{\ast} = 10^7 \erg \g^{-1}$.
The essential point is that collisions between comparably sized grains
in the AU Mic disk are destructive.\footnote{Commercial
sandblasting machines accelerate abrasive particles
up to speeds of 100 m s$^{-1}$.}

%
%

\subsection{Blow-out by Stellar Wind and Radiation Pressure}
\label{blowout}

Grains of certain sizes cannot occupy
orbits bound to the star because of stellar
wind and radiation (SWR) forces.
The ratio of repulsive to gravitational forces felt by a grain equals

\begin{eqnarray}
\beta & = &\frac{F_{\rm rad} + F_{\rm wind}}{F_{\rm grav}} \\
 & = & \frac{3}{16\pi} \frac{L_{\ast} P_{\rm SWR}}{GM_{\ast}c\rho s}\,,
\end{eqnarray}

\ni where the dimensionless factor

\begin{equation}
P_{\rm SWR} \equiv Q_{\rm rad} + Q_{\rm wind} \frac{\dot{M}_{\ast} v_{\rm wind} c}{L_{\ast}}
\end{equation}

\ni measures the extent to which the ram pressure exerted by the
(assumed radial) wind dominates radiation pressure.\footnote{The
assumption of a purely radial wind is valid insofar as
the azimuthal velocity of the wind $v_{{\rm wind},\phi}$
is less than $\Omega r$.
By modelling the stellar magnetic field
as that of a rotating (split)
monopole \citep{weberdavis,parker},
we estimate that $v_{{\rm wind},\phi}/\Omega r \lesssim 1/30$.}
Here $s$ and $\rho \sim 2 \gm \cm^{-3}$
are, respectively,
the radius and internal density of a particle, $G$ is the gravitational
constant, $c$ is the speed of light,
$Q_{\rm rad} \lesssim 2$ is the cross section that the grain presents
to radiation pressure in units of the geometric cross section \citep{bls},
and $Q_{\rm wind} \sim 1$ is the analogous dimensionless
cross section the grain presents to wind pressure.\footnote{Our
$Q_{\rm rad}$ equals $Q_{\rm pr}$ from \citet{bls}.
It should not be confused with $Q_{\rm scat}$, the
usual scattering efficiency.}
For $Q_{\rm rad} \sim 2$ (appropriate for the $s > \lambda_{\ast}/2\pi$
geometric optics limit where $\lambda_{\ast} \approx 1 \mum$
is the wavelength at which the bulk of the stellar luminosity
is emitted),
wind pressure is negligible compared to radiation pressure except
for the largest value of $\dot{M}_{\ast}$ considered.
Table \ref{windtable} lists possible values for $P_{\rm SWR}$.

Grains are continually created from colliding parent bodies.
Throughout this paper, we assume that
parent bodies move on nearly circular ($\beta \ll 1$)
orbits, and that the velocities with which grains are ejected from
parent bodies are
small compared to parent body orbital velocities.
These assumptions imply that
grains having $\beta \geq 1/2$ are ``blown out'' by SWR pressure.
For constant $P_{\rm SWR}$ with $s$, the condition
$\beta \geq 1/2$ is equivalent to

\begin{equation}
s \leq s_{\rm blow} = \frac{3}{8\pi} \frac{L_{\ast} P_{\rm
    SWR}}{GM_{\ast}c\rho} \sim 0.2 \left( \frac{P_{\rm SWR}}{2}
\right) \mum \,.
\label{sblow}
\end{equation}

\ni Grains for which $s < s_{\rm blow}$ are unbound and move on hyperbolic
escape trajectories. Grains for which $s = s_{\rm blow}$
move on parabolic escape trajectories.
A grain for which $s - s_{\rm blow} \ll s_{\rm blow}$ moves initially on a
highly elliptical orbit whose periastron distance $r_{{\rm peri},0}$
equals the orbital radius of the parent body.
It is these ``barely bound'' grains
that contribute significantly to the scattered light
observed in the outer disk. The initial eccentricity $e_0$ and
semi-major axis $a_0$
of a barely bound grain upon its birth
are uniquely related to grain size $s$ via the force ratio $\beta$:

\begin{subequations}\label{e0}
\begin{equation}
e_0 = \frac{\beta}{1-\beta} \, ,
\end{equation}
\begin{equation}\label{a0_cpr}
a_0 = \frac{r_{{\rm peri},0}}{1-e_0} \, ,
\end{equation}
\begin{equation}
\beta = \frac{1}{2} \left( \frac{s_{\rm blow}}{s} \right) \,,
\end{equation}
\end{subequations} 

\ni where the last relation assumes that $P_{\rm SWR}$ is independent of $s$.

A consequence of equations (\ref{tcol}) and (\ref{e0}) is that
grains on highly eccentric orbits have prolonged collision times.
Provided that the optical depth traversed by a grain is concentrated near
periastron at $r = r_{\rm peri}$, the optical depth
$\tau_{\perp}$ in (\ref{tcol})
should be evaluated at $r = r_{\rm peri}$.
However, $\Omega$ should not necessarily be evaluated
for a semi-major axis equal to $r_{{\rm peri}}$.
Instead, from (\ref{a0_cpr}),

\begin{equation}\label{Omega}
\Omega(e) = \left( \frac{GM_{\ast}}{r_{\rm peri}^3} \right)^{1/2} (1-e)^{3/2} \, .
\end{equation}

\ni 
A related useful quantity is the fraction of time a grain on a fixed orbit
spends at radii between $r_{\rm peri}$ and $r_{\rm peri} + \Delta r$:

\begin{equation}\label{f(e)}
f(e) \sim \left( \frac{\Delta r}{r_{\rm peri}} \right)^{1/2}
\frac{(1-e)^{3/2}}{(1+e)^{1/2}} \,,
\end{equation}

\ni valid in the limit of $\Delta r \ll r_{\rm peri}$ and large $e$.

\subsection{Corpuscular and Poynting-Robertson Drag}
\label{cprdrag}

Under the drag due to corpuscular and Poynting-Robertson (CPR) forces
\citep[see, e.g.,][]{bls},
dust grain orbits having periastron distances $r_{\rm peri}$ and
arbitrary eccentricities $e$ collapse to a point in a time

\begin{equation}
t_{\rm CPR} = 
\frac{4\pi c^2 \rho}{3L_{\ast}P_{\rm CPR}} E(e) r_{\rm peri}^2 s \,,
\label{tcpr}
\end{equation}

\ni where the dimensionless factor

\begin{equation}
P_{\rm CPR} \equiv Q_{\rm rad} + 
Q_{\rm wind} \frac{\dot{M}_{\ast} c^2}{L_{\ast}}
\end{equation}

\ni quantifies the relative importance of wind-driven
versus radiation-driven drag. 
For the values of $\dot{M}_{\ast}$ we consider, the stellar wind is at least
as important as the stellar radiation:
$P_{\rm CPR} > 2$; see Table \ref{windtable}.  
The dimensionless factor

\begin{equation}\label{WW_Edef}
E(e) = \frac{8}{5} \frac{(1+e)^2}{e^{8/5}} 
\int_0^{e} \frac{x^{3/5} \, dx}{(1-x^2)^{3/2}}
\end{equation}

\ni governs the decay of orbital eccentricity \citep{wyattwhipple}.
A related useful result from \citet{wyattwhipple} is that

\begin{equation}\label{dedt}
\frac{de}{dt} 
= \left(\frac{15L_\ast P_{\rm CPR}}{32\pi c^2 \rho r_{\rm peri,0}^2} \right)
\frac{1}{s} \frac{e_0^{8/5}}{(1+e_0)^{2}}
\frac{(1-e^2)^{3/2}}{e^{3/5}}
\, .
\end{equation}

The initial effect of
CPR drag on a highly eccentric orbit is to reduce the apastron distance
while keeping the periastron distance and eccentricity nearly fixed 
\citep[see Figure 1 of][]{wyattwhipple}.
For highly eccentric orbits,
the time spent during this apastron reduction
phase is much longer than the usual CPR-timescales
that are cited for $e \ll 1$.
As $e$ approaches 1, $E$ diverges as 

\begin{equation}\label{Egrowth}
E(e \approx 1) \propto (1-e)^{-1/2} \,.
\end{equation}

\ni Once CPR drag reduces
the eccentricity to values less than a few percent, the entire orbit
collapses to a point in a time given by $t_{\rm CPR}$ with $E \approx 1$.

\begin{figure}
\epsscale{1.3}
\hspace{-.9cm}
\plotone{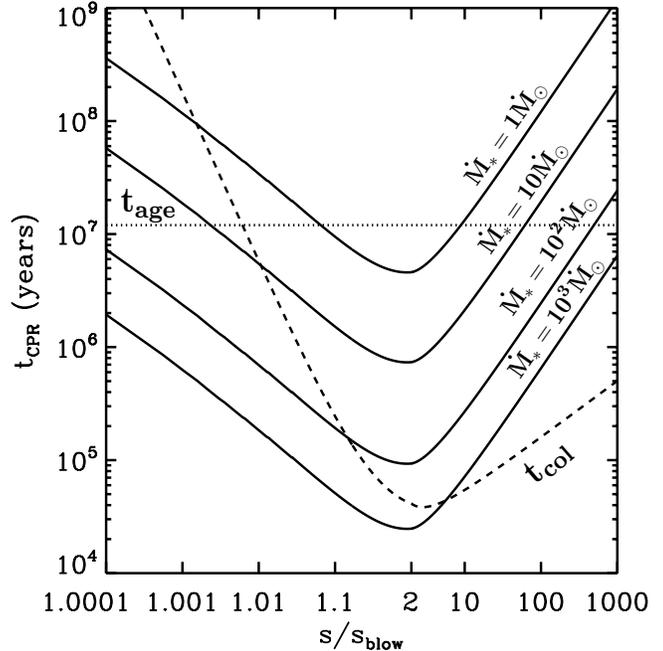}
\caption{CPR drag time as a function of grain size $s/s_{\rm blow}$, for
  $\dot{M}_{\ast}/\dot{M}_{\odot} = \{1, 10, 10^2, 10^3\}$.  To
  highlight the behavior as $s\rightarrow s_{\rm blow}$, the
  horizontal axis is scaled as $\log(s/s_{\rm blow}-1)$.
  The fiducial collision
  time $t_{\rm col} = t_{{\rm c, BR}}(s/s_{\rm blow})^{1/2}(1-e_0)^{-3/2}$ 
  is also indicated as a dashed line (see \S\ref{realtcol}).  
  The
  size $s_{\rm break}$ corresponds approximately to where this
  collision time and
  the CPR drag time are equal.
  Timescales for removal by CPR drag and collisions
  can be much shorter than the age of the system, $t_{\rm age}$.
  \label{tcprfig}}
\end{figure}

Table \ref{windtable} provides sample values of $t_{\rm CPR} (e =
e_0)$ for three choices
of $s/s_{\rm blow} = \{1.1,1.5,15\}$, while Figure \ref{tcprfig} depicts how
$t_{\rm CPR} (e = e_0)$ varies with $s$, all for 
$r_{\rm peri} = r_{\rm BR} = 43\AU$.

\section{THEORY}
\label{theory}

We assemble the ingredients laid out in \S\ref{oom}
into a theory for the distribution of grain sizes (\S\ref{cascades})
and the profile of optical depth (\S\ref{physics})
in the AU Mic disk.
Included in our analysis are
estimates of the sizes and total mass of the largest parent bodies
undergoing
a collisional cascade.
According to our theory, all of the current optical to near-IR
observations probe grains whose population is maintained in steady
state and that are still bound---most only barely---to
the central star. Contributions to scattered light from unbound
and unequilibrated populations of grains
are assessed in \S\ref{unbound} and \S\ref{sage}, respectively.

\subsection{Equilibrium Size Distribution} 
\label{cascades}
We posit an annulus of parent bodies extending from 
$r = r_{\rm BR}-\Delta r/2$ 
to $r=r_{\rm BR}+\Delta r/2$---the
``birth ring''---where grains are born from the collisional attrition
of larger parent bodies.  
These grains travel on orbits whose eccentricities and semi-major
axes are continuously modified by CPR drag. Grains are removed from 
the birth ring
either by CPR drag or by collisions with other grains.
The goal of this section is to determine the equilibrium size
distribution $dN/ds$ as a function of $r$, where $dN$ is the vertical
column density of grains having sizes between $s$ and $s+ds$.

The size distribution of dust in debris disks is commonly assumed
to be proportional to $s^{-q_{\rm ce}} = s^{-7/2}$.
This scaling is appropriate for grains on low-eccentricity orbits
whose collisional strengths are independent of size, whose
spatial distribution is homogeneous,
and whose numbers are maintained
in a purely collisional equilibrium, as first derived by
Dohnanyi (1969; see also O'Brien \& Greenberg 2003; Pan \& Sari 2005).
In such equilibrium cascades, as much mass is ground
into every size bin as is ground
out.  One presumption behind the cascade is that collision
times are short
enough that the system has relaxed into collisional equilibrium.
However, Figure \ref{tcprfig}
shows that $t_{\rm CPR}$ and $t_{{\rm col}}$ are of the same order for
some grain sizes.
Visible dust may be removed
too quickly by CPR drag to participate in a purely collisional, equilibrium
cascade.  




Instead of making the usual assumption that
the size distribution is proportional to $s^{-7/2}$ for all $s$, we
construct the
following model.
We define $s_{\rm break}$ as the radius of the grain for which

\begin{equation}
t_{\rm col} (s_{\rm break}) = t_{\rm CPR} (s_{\rm break})
\label{sbreak}
\end{equation}

\ni at $r=r_{\rm BR}$.
We expect that grains of a given size can participate in a standard
collisional cascade if they have had enough time to collide destructively
about once
(for numerical estimates of the time required for a cascade to equilibrate
see, e.g., Campo Bagatin et al.~1994 and references therein).
For $s > s_{\rm break}$, we expect $t_{\rm CPR} > t_{\rm col}$ and
a Dohnanyi-like size distribution (modified appropriately for grains
on highly eccentric orbits, i.e., for spatial inhomogeneity).
For $s < s_{\rm break}$,
we will see that $t_{\rm CPR} < t_{\rm col}$.
The two regimes are treated in \S\ref{morebreak} and \S\ref{eqsize},
respectively.
We discuss which grain sizes carry the bulk of the optical depth and
how collision times vary with grain size in \S\ref{realtcol}.
These considerations are applied
to computing $s_{\rm break}$ in \S\ref{fullopti}. The sizes of the largest
parent bodies are estimated in \S\ref{maxcolsize}.

Readers interested only in our results for the grain size
distribution can examine equations 
(\ref{dNtilde-pre}),
(\ref{theunusual}), 
(\ref{dnds_fe}), 
(\ref{theusual}),
(\ref{dohn+e}),
(\ref{dohn+e1}),
(\ref{sbreak_value}),
and (\ref{smax_value}),
and can skip ahead to \S\ref{physics}
where grain dynamics are analyzed.

\subsubsection{Equilibrium Size Distribution for $s < s_{\rm break}$}
\label{eqsize}

First we define 

\begin{equation}\label{dNtilde-pre}
\frac{d\mathcal{N}}{ds} \equiv 
\int_{0}^{\infty} \frac{dN}{ds}\, 2\pi r \, dr \,  
\end{equation}

\ni as the size distribution of grains integrated over the entire disk.



Fresh debris having $s < s_{\rm break}$ continually sprays
from colliding bodies having $s > s_{\rm break}$.
We assume that the initial or ``injection'' spectrum of fresh debris
follows a power law;
the rate at which grains having sizes between $s$ and $s+ds$
are injected into the entire disk obeys

\begin{equation}\label{inject}
\left.\frac{d\dot{\mathcal{N}}}{ds}\right|_{\rm I} = C s^{-q_{\rm inject}} \,,
\end{equation}

\ni where $C$ is a constant and the subscript ``I'' denotes ``injection.''
Theoretical considerations
of mass conservation suggest $q_{\rm inject} = 3$--4 \citep{greenberg},
while impact experiments using centimeter-sized targets
suggest values of $q_{\rm inject} \approx 3.5$--4
\citep[see their Figure 3]{fujiwara}.

To solve for the steady-state size distribution, we
equate the injection rate $d\dot{\mathcal{N}}/ds|_{\rm I}$ to the removal rate
$d\dot{\mathcal{N}}/ds|_{\rm R}$.
We show in \S\ref{realtcol} that removal
is dominated by CPR drag onto the central star:
$t_{\rm CPR} / t_{\rm col} < 1$ for $s < s_{\rm break}$.
Then grains having $s < s_{\rm break}$
drag inward from the birth ring largely unimpeded by collisions,
and

\begin{equation}
\left.\frac{d\dot{\mathcal{N}}}{ds}\right|_{\rm R}
= \frac{1}{t_{\rm CPR}(s)} \frac{d\mathcal{N}}{ds} \, .
\end{equation}

\ni 
Equating the injection and removal rates
yields 

\begin{equation}
\frac{d\mathcal{N}}{ds}
\sim C s^{-q_{\rm inject}} t_{\rm CPR}(s,
r_{\rm peri} =
r_{\rm BR})\,\,\,\,\,\, {\rm for} \,\,\,\, s < s_{\rm break} \,.
\label{theunusual}
\end{equation}



The column density $dN/ds$ local to the birth ring is proportional
to $d\mathcal{N}/ds$ times the fraction of time spent inside the birth
ring:

\begin{eqnarray}
\left. \frac{dN}{ds} \right|_{\rm BR} & 
\propto & \frac{d\mathcal{N}}{ds} 
\frac{\left.\Delta t_{\rm reside}\right|_{\rm CPR}}{t_{\rm CPR}} \\
 & \propto & s^{-q_{\rm inject}} \left.\Delta t_{\rm
 reside}\right|_{\rm CPR}
\,\,\,\, {\rm for} \,\,\, s < s_{\rm break} \,,
\label{dnds_fe}
\end{eqnarray}

\begin{figure}
\epsscale{1.4}
\hspace{-.7in}
\plotone{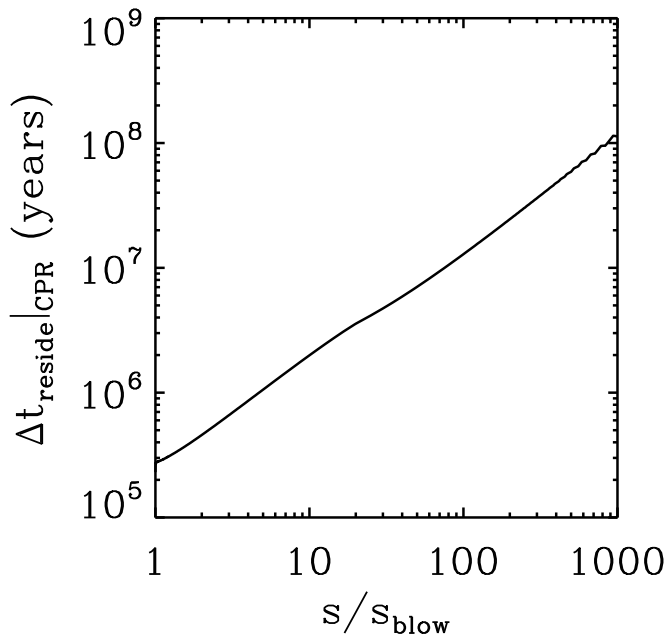}
\caption{The length of time $\Delta t_{\rm reside}|_{\rm CPR}$ a grain
    spends within the birth ring (at radii between $r_{\rm BR}-\Delta
    r/2$ and $r_{\rm BR}+\Delta r/2$)
    as a function of grain size, for grains whose lifetimes are limited by
    CPR drag.  
We take $r_{\rm peri,0} = r_{\rm BR} = 43\AU$,
    $\Delta r/r_{\rm BR} = 0.1$, and $\dot{M}_\ast = 1\dot{M}_\odot$.
  \label{residefig}}
\end{figure}

\ni where $\Delta t_{\rm reside}|_{\rm CPR}$ is the total time
a grain spends inside the birth ring over its CPR-limited lifetime.
Figure \ref{residefig} plots $\Delta t_{\rm reside}|_{\rm CPR}$
as a function of $s$ for the parameters $\dot{M}_{\ast}/\dot{M}_{\odot} = 1$,
$\Delta r / r_{\rm BR} = 0.1$, and $r_{\rm peri,0} = r_{\rm BR}$.
We compute this quantity by explicitly tracking the position of a
grain on a decaying orbit.
Evidently, $\Delta t_{\rm reside}|_{\rm CPR}$ scales approximately
linearly with $s$ for all $s$. We can understand this linear scaling
by examining two extremes. For $s \gg s_{\rm blow}$, orbits are nearly
circular always, and $\Delta t_{\rm reside}|_{\rm CPR}$ is merely
the time for the grain's orbital radius to decay by $\Delta r / 2$.
This time is proportional to $s$ since $E \approx 1$ (see equation
[\ref{tcpr}]). For $s - s_{\rm blow} \ll s_{\rm blow}$, eccentricities
$e_0$ are large.
Over most of a grain's lifetime, the grain's periastron lies inside the birth
ring while its apastron lies well
outside. Evaluated over intervals shorter than $t_{\rm CPR}$,
the fraction of time
the grain spends inside the birth ring is given by $f(e)$ with $r_{\rm peri}$
set equal to $r_{\rm BR}$. Applying (\ref{f(e)}) and (\ref{dedt}), we have

\begin{eqnarray}
\left.\Delta t_{\rm reside}\right|_{\rm CPR}
& \sim & \int_{e_0}^0 f(e) \frac{dt}{de}\,de \nonumber \\
& \propto & s \left( \frac{\Delta r}{r_{\rm BR}} \right)^{1/2}
\int_0^{e_0} \left(\frac{1+e_0}{1+e}\right)^2
\left(\frac{e}{e_0}\right)^{3/5}\frac{de}{e_0} \nonumber \\ & \propto & s \, ,
\end{eqnarray}

\ni where the integral in the second row is nearly constant with $s$.


\subsubsection{Equilibrium Size Distribution for $s > s_{\rm break}$}
\label{morebreak}

For $s > s_{\rm break}$, a collisional cascade
is established before CPR drag has time to remove grains. Collisions
occur primarily in the birth ring since the vertical optical depth
is greatest there. In the birth ring, the usual collisional
equilibrium implies

\begin{equation}
\left. \frac{dN}{ds} \right|_{\rm BR}
\propto s^{-q_{\rm ce}} 
= s^{-7/2} \,\,\,\, {\rm for} \,\,\, s > s_{\rm break} \,.
\label{theusual}
\end{equation}


\ni 
By the same logic that led to (\ref{dnds_fe}),

\begin{equation}
\frac{d\mathcal{N}}{ds} 
 \propto 
\left. \frac{dN}{ds} \right|_{\rm BR} \frac{t_{\rm col}}{\left.\Delta t_{\rm
    reside}\right|_{\rm col}}
\,\,\,\, {\rm for} \,\,\, s > s_{\rm break} \,,
\label{dohn+e}
\end{equation}

\ni where we have assumed (and show in \S\ref{realtcol}) that
$t_{\rm col} < t_{\rm CPR}$ is the appropriate lifetime for grains
having $s > s_{\rm break}$. By analogy to $\Delta t_{\rm reside}|_{\rm CPR}$,
$\Delta t_{\rm reside}|_{\rm col}$ is the total time a grain spends within
the birth ring over a collision-limited lifetime.

In the special case of large initial eccentricity $e_0$,
\begin{eqnarray}
\frac{d\mathcal{N}}{ds} 
& \propto &
\left. \frac{dN}{ds} \right|_{\rm BR} \frac{1}{f(e_0)} \nonumber \\
& \propto & \left. \frac{dN}{ds} \right|_{\rm BR} (1 - e_0)^{-3/2} \,,
 \,\, s > s_{\rm break}, \, e_0 \approx 1 \,,
\label{dohn+e1}
\end{eqnarray}

\ni where we have used the fact that grains having large $e_0$ and whose
lifetimes are limited by $t_{\rm col} < t_{\rm CPR}$ have their periastra
within the birth ring and their eccentricities close to their birth values
for nearly all their lives.


\subsubsection{Optical Depth and Collision Times in the Birth Ring}
\label{realtcol}

We estimate the sizes of grains that carry the lion's share of
the optical
depth in the birth ring.  For $s > s_{\rm break}$, the column density
in the birth ring obeys 
$dN/ds \propto s^{-7/2}$.
The vertical optical depth
contributed by such grains equals

\begin{equation}\label{taugtbreak}
\tau_\perp(s>s_{\rm break}) \sim s^3\frac{dN}{ds} \propto s^{-1/2} \, ;
\end{equation}

\ni therefore among grains of size $s > s_{\rm break}$, the optical depth
is dominated by grains of size $s_{\rm break}$.
What about the regime $s < s_{\rm break}$?
From (\ref{dnds_fe}), 

\begin{equation}\label{taultbreak}
\tau_\perp(s<s_{\rm break}) \sim
s^3\frac{dN}{ds}
\propto s^{3-q_{\rm inject}} \left. \Delta t_{\rm reside} \right|_{\rm CPR}
\, .
\end{equation}

\ni For $q_{\rm inject} \approx 3.5$--4, this quantity either
grows or is approximately
constant with $s$. 
We conclude that grains of size $s \sim s_{\rm break}$ dominate
the total geometric optical depth in the birth ring:

\begin{equation}
\tau_{\perp}(s_{\rm break}) \sim \tau_{\perp,{\rm BR}} \, ,
\label{everuseful}
\end{equation}

\ni which in combination with (\ref{tcolvis}), (\ref{sproj}) and
(\ref{Omega}) implies that

\begin{equation}
t_{\rm col} (s_{\rm break}) \sim t_{\rm c,BR}(1-e_{0,\rm break})^{-3/2} \,,
\label{everuseful2}
\end{equation}

\ni where $e_{0,{\rm break}} \equiv e_0 (s_{\rm break})$.

We exploit (\ref{tcolvis}), (\ref{Omega}), and
(\ref{taugtbreak})--(\ref{everuseful2})
to estimate $t_{\rm col}$ for arbitrary $s$.
For $s > s_{\rm break}$, 

\begin{equation}\label{tcolfull}
t_{\rm col}(s>s_{\rm break}) \sim t_{{\rm c,BR}} \left(\frac{s}{s_{\rm
    break}}\right)^{1/2} (1-e_0)^{-3/2} \, ,
\end{equation}

\ni where we have set $e=e_0$ since most of the grain's lifetime is
 spent with that eccentricity (see \S\ref{cprdrag}).
 Since in the large-$s$ limit
$t_{\rm col} \propto s^{1/2}$ while $t_{\rm CPR} \propto s$,
the assumption made in \S\ref{morebreak} that grains are removed
principally by collisions
for $s > s_{\rm break}$ is asymptotically valid.
In Figure \ref{tcprfig}, we plot (\ref{tcolfull})
as a fiducial size-dependent
collision time, replacing
$s_{\rm break}$ by
$s_{\rm blow}$ to render the curve independent of stellar
mass-loss rate. This replacement, performed solely for Figure \ref{tcprfig},
accrues only a slight error since we find in \S\ref{fullopti} that
$s_{\rm break}$ and $s_{\rm blow}$ are nearly the same.


Next we estimate $t_{\rm col}(s < s_{\rm break})$ and
show that $t_{\rm CPR} / t_{\rm col} < 1$ for $s < s_{\rm break}$, as was
assumed in \S\ref{eqsize}.
Equation (\ref{everuseful}) implies that 
grains having $s < s_{\rm break}$ are predominantly destroyed 
by $s_{\rm break}$-sized grains.  Then\footnote{Equation
  (\ref{tcollessbreak}) overestimates $t_{\rm col}$ because it neglects
  the fact that grains on highly eccentric orbits intercept an optical
  depth parallel to the disk in addition to $\tau_{\perp,{\rm BR}}$.
  This neglect does not change our derived scaling relations, but it
  will change certain normalizations, e.g., the threshold
  $\dot{M}_\ast$ dividing CPR-dominated from collision-dominated
  behavior.  We are indebted to Y. Wu for pointing out this omission,
  which will need to be corrected in future work.}

\begin{eqnarray}\label{tcollessbreak}
t_{\rm col} (s < s_{\rm break}) 
& \sim & \frac{1}{\Omega(s) \tau_{\perp}(s_{\rm break})} \nonumber \\
& \sim & t_{\rm col} (s_{\rm break}) 
\left( \frac{1-e_{0,{\rm break}}}{1-e_0} \right)^{3/2} \,.
\end{eqnarray}

\ni For convenience, we construct the
approximate fitting formula for $t_{\rm CPR}$ from (\ref{tcpr}),
(\ref{Egrowth}), and (\ref{sbreak}):

\begin{equation}\label{fitting}
t_{\rm CPR} (s < s_{\rm break}) \sim t_{\rm col} (s_{\rm break}) 
\left( \frac{s}{s_{\rm break}} \right) 
\left( \frac{1-e_{0,{\rm break}}}{1-e_0} \right)^{1/2} \,.
\end{equation}

\ni Combining (\ref{tcollessbreak}) and (\ref{fitting}), we find that 
the ratio between CPR and collision lifetimes is

\begin{equation}
\left. \frac{t_{\rm CPR}}{t_{\rm col}} \right|_{s < s_{\rm break}}
\sim \left( \frac{s}{s_{\rm break}} \right) \left(
\frac{1-e_0}{1-e_{0,{\rm break}}} \right) < 1 \,;
\label{theend}
\end{equation}

\ni indeed this ratio vanishes as $s$ approaches $s_{\rm blow}$. We have
established that inequality (\ref{theend}) holds at $r = r_{\rm BR}$,
but in fact it holds for all $r = r_{\rm peri} < r_{\rm BR}$, since
$t_{\rm CPR} \propto r^2$ while $t_{\rm col} \propto r^{3/2}$.
Therefore our
assumption that bound grains having $s < s_{\rm break}$
are removed principally by CPR drag is valid.

\subsubsection{Calculating $s_{\rm break}$}
\label{fullopti}

Equate the collision time (\ref{tcolfull}) to the
CPR drag time (\ref{tcpr}),

\begin{equation}
(1-e_{0,{\rm break}})^{-3/2} t_{{\rm c,BR}}
\sim \frac{4 \pi c^2 \rho r_{\rm BR}^2}{3 L_{\ast} P_{\rm CPR}} s_{\rm break} 
E(e_{0,{\rm break}})\,,
\end{equation}

\ni to find 

\begin{equation}\label{sbreak_value}
s_{\rm break} \sim \left\{ \begin{array}{lll}
1.002s_{\rm blow} & = 0.2\mum &  
\,\,\,\, {\rm for} \,\,\, \dot{M}_{\ast}/\dot{M}_{\odot} = 1 \\
1.01s_{\rm blow} & = 0.2\mum &  
\,\,\,\, {\rm for} \,\,\, \dot{M}_{\ast}/\dot{M}_{\odot} = 10 \\
1.1s_{\rm blow} & = 0.3\mum & 
 \,\,\,\, {\rm for} \,\,\, \dot{M}_{\ast}/\dot{M}_{\odot} = 10^2 \\
2.3s_{\rm blow} & = 2\mum & 
 \,\,\,\, {\rm for} \,\,\, \dot{M}_{\ast}/\dot{M}_{\odot} = 10^3 \, .
\end{array} \right. 
\end{equation}

\subsubsection{Larger Parent Bodies}
\label{maxcolsize}

Bodies having $s > s_{\rm break}$ participate in a
collisional cascade in which the Dohnanyi-like spectrum extends from
$s_{\rm break}$ up to a maximum size $s_{\rm max}$. By definition,
$s_{\rm max}$ characterizes those grains
whose collisional lifetimes equal the age of the AU Mic disk; by
(\ref{tcolfull}), this size satisfies

\begin{equation}
t_{\rm col} (s_{\rm max}) \sim t_{{\rm c,BR}}
\left(\frac{s_{\rm max}}{s_{\rm break}}\right)^{1/2} = t_{\rm age} \,,
\end{equation}

\ni where we have dropped the eccentricity-dependent factor 
since grains having $s= s_{\rm max} \gg s_{\rm blow}$ 
travel on essentially circular orbits.  
Then 

\begin{equation}\label{smax_value}
s_{\rm max} \sim \left(\frac{t_{\rm age}}{t_{{\rm c,BR}}}\right)^2
s_{\rm break} 
\sim \left\{ \begin{array}{rl}
10\cm &  \,\,\,\, {\rm for} \,\,\, \dot{M}_{\ast}/\dot{M}_{\odot} =
1 \\
10\cm &  \,\,\,\, {\rm for} \,\,\, \dot{M}_{\ast}/\dot{M}_{\odot} =
10 \\
20\cm &  \,\,\,\, {\rm for} \,\,\,
\dot{M}_{\ast}/\dot{M}_{\odot} = 10^2 \\ 
100\cm &  \,\,\,\, {\rm for} \,\,\,
\dot{M}_{\ast}/\dot{M}_{\odot} = 10^3 \,.
\end{array} \right.
\end{equation}

Our model has no information on parent bodies having $s > s_{\rm max}$.
While such bodies likely exist, we do not know whether they are
currently in a constructive (planet building) or destructive
(debris generating)
phase of their evolution.  It is not justified to extend the size
distribution to $s > s_{\rm max}$ (e.g., to the kilometer size range)
as is sometimes done.

\subsection{Physical Implications of Optical Depth Profiles}
\label{physics}

The birth ring divides the inner disk from the outer disk.
In \S\ref{innerphysics}, we discuss the
vertical optical depth in the inner disk.
In \S\ref{totalmass}, we estimate the total disk mass.
In \S\ref{outerphysics}, we derive
analytically how the vertical
optical depth scales with radius in the outer disk.

\subsubsection{Inner Disk $(r < r_{\rm BR})$: Competition Between Collisions
  and CPR-Driven Accretion}
\label{innerphysics}

We showed in \S\ref{realtcol} that grains of size
$s \sim s_{\rm break}$ make the largest contribution
to the total optical depth at $r\approx r_{\rm BR}$.
Bound grains having $s<s_{\rm break}$ tend to be transported inwards
by CPR drag, unimpeded by interparticle collisions.  Larger
grains tend to be collisionally destroyed before reaching the star.
How does the vertical optical depth in the inner disk compare with the
optical depth in the birth ring?

We define CPR-dominated type A disks to be systems for which
$s_{\rm break}-s_{\rm blow} \gg s_{\rm blow}$.  In such disks, 
grains for which $s_{\rm blow} < s < s_{\rm break}$ are numerous, contain
a significant fraction of the total optical depth in the birth ring,
and tend to accrete onto the central star before undergoing a collision.
From continuity, the optical depth $\tau_\perp(r<r_{\rm BR})$ scales
approximately as $r^0$: the inner disk is ``filled in.''

By contrast, in collision-dominated type B disks, 
$s_{\rm break}-s_{\rm blow} \ll s_{\rm blow}$.
The region inside the birth ring is virtually empty.
The reasons for this are twofold.
First, the range of sizes of grains that drag in without being
collisionally destroyed ($s_{\rm blow} < s < s_{\rm break}$) is narrow;
there are not many such grains.
Second, because $s_{\rm break}$ is so close to $s_{\rm blow}$, 
$s_{\rm break}$-sized grains have large initial eccentricity.
They spend most of their lifetimes having
$r_{\rm peri} \approx r_{{\rm peri},0}$ and only a small portion at
$r < r_{\rm BR}$ (see \S\ref{cprdrag}).

As judged from (\ref{sbreak_value}), if $\dot{M}_\ast/\dot{M}_\odot \gg
10^2$, type A conditions would hold for the AU Mic disk and the inner
disk would be filled in.  If
$\dot{M}_\ast/\dot{M}_\odot \ll 10^2$, then type B conditions would hold
and the inner disk would be empty. In \S\ref{montecarlo}, we not
only check these assertions by detailed Monte Carlo simulations of the
AU Mic disk, but also decide which case is favored by the observations.

\subsubsection{Total Mass of the Disk}
\label{totalmass}

By (\ref{theusual}), most of the mass of the disk is contributed by
the largest grains ($s=s_{\rm max}$), since $s^4dN/ds \propto s^{1/2}$. 
We scale from the column density of $s_{\rm break}$-sized grains in the
birth ring,

\begin{equation}
\tau_{\perp,{\rm BR}} \sim \pi s_{\rm break}^3 \left.\frac{dN}{ds}
\right|_{s_{\rm break}}
\end{equation}

\ni to estimate the column density of $s_{\rm max}$-sized grains in the birth
ring,

\begin{equation}
s_{\rm max}
\left.\frac{dN}{ds}\right|_{s_{\rm max}}
\sim \frac{\tau_{\perp,{\rm BR}}}{\pi
  s_{\rm break}^3}\left(\frac{s_{\rm max}}{s_{\rm
  break}}\right)^{-7/2}s_{\rm max} \, .
\end{equation}

\ni The total number of such grains is their column density multiplied
by the area of the birth ring $2\pi r_{\rm BR} \Delta r$, 
since $s_{\rm max}$-sized grains
undergo a collision long before dragging 
 inwards.
Multiplying this total number by the mass of a single grain yields the
mass of the disk,

\begin{eqnarray}\label{parentmass}
M_{\rm max} & \sim & \frac{8\pi}{3} \rho r_{\rm BR} 
\tau_{\perp,{\rm BR}} s_{\rm break}
\left(\frac{s_{\rm max}}{s_{\rm break}}\right)^{1/2}\Delta r \\
&  \sim &
0.01M_{\oplus} \left(\frac{\Delta r/r_{\rm BR}}{0.1}\right) 
\left(\frac{\tau_{\perp,{\rm BR}}}{4\times10^{-3}}\right)^2
\left(\frac{s_{\rm break}}
{0.2 \mum}\right) \nonumber \, ,
\end{eqnarray}

\ni where we have used (\ref{tcolvis}) and (\ref{smax_value}),
and have normalized $\Delta r / r_{\rm BR}$ and $\tau_{\perp,{\rm BR}}$
to values that yield good fits to observations, as described
in \S\ref{montecarlo}.
The steady comminution of $\sim$0.01$M_{\oplus} \sim$ 1 lunar mass's
worth of decimeter-sized bodies into micron-sized particles does not
seem an unduly heavy burden for the AU Mic system to bear.
The solar system is thought to have somehow shed $\sim$$10 M_{\oplus}$
of rock and ice near $\sim$30 AU over an uncertain
timescale of $10$--$1000$ Myr during its ``clean-up'' phase
\citep{gls}.

\subsubsection{Outer Disk $(r > r_{\rm BR})$: Barely Bound Grains}
\label{outerphysics}

Grains created at $r = r_{\rm BR} = 43 \AU$ and having 
$s - s_{\rm blow} \lesssim s_{\rm blow}$ occupy initially highly eccentric
orbits having periastron distances $r_{\rm peri} = r_{\rm BR}$ (see \S\ref{blowout}).
Here we show that such barely bound grains establish an outer disk at
$r \gg r_{\rm BR}$ whose vertical optical depth scales approximately as $r^{-5/2}$
for CPR-dominated type A disks, and as $r^{-3/2}$
for collision-dominated type B disks.
The contribution
of unbound grains having
$s \le s_{\rm blow}$ relative to that of barely bound
grains is assessed in \S\ref{unbound}.

The outer disk comprises grains
having sizes slightly greater than
$s_{\rm blow}$ since only those grains have
substantial eccentricities (see equations [\ref{e0}]).
We refer to such barely bound grains as having sizes $s = s_{\rm blow,+}$.
The optical depth in the outer disk should scale
approximately as their column density $N_{{\rm blow,+}}$:

\begin{equation}
\tau_\perp \propto N_{\rm blow,+} \propto \frac{1}{r}\frac{d\mathcal{N}_{\rm blow,+}}{dr} \, ,
\label{massnumber}
\end{equation}

\ni where $\mathcal{N}_{\rm blow,+}$ is the total number of
such grains in the entire disk (see
    [\ref{dNtilde-pre}]).
By the chain rule,



\begin{equation}
\frac{d\mathcal{N}_{\rm blow,+}}{dr} = \frac{d\mathcal{N}_{\rm blow,+}}{de} \frac{de}{dr} \,.
\label{chain}
\end{equation}

\ni
We determine $de/dr$ by making the approximation that at any instant,
all grains are located at their apastra:
%

\begin{eqnarray}
e & = & \frac{r_{\rm apo} - r_{\rm peri}}{r_{\rm apo} + r_{\rm peri}} \\
  & \approx & \frac{r - r_{\rm BR}}{r+r_{\rm BR}} \,. \label{approxapo}
\end{eqnarray}

\ni This approximation should be good for the highly eccentric
orbits of the outer disk.
\ni Differentiating (\ref{approxapo}), we find for $r \gg r_{\rm BR}$ that

\begin{equation} \label{dedr}
\frac{de}{dr} \approx \frac{2r_{\rm BR}}{(r+r_{\rm BR})^2} \propto \frac{1}{r^2} \,.
\end{equation}

The remaining factor in (\ref{chain}), $d\mathcal{N}_{\rm blow,+}/de$,
is determined by the size distribution of barely bound grains. This
differs between type A and type B disks, as seen below.
The size distribution determines the initial shape of the eccentricity
distribution. The eccentricity distribution is altered over time by CPR
drag. Consider grains
having identical initial eccentricities $e_0$
created at a constant rate $\dot{\widetilde{\mathcal{N}}}$.
In steady state, CPR drag transports a constant number of particles
per time through eccentricity space, $(d\widetilde{\mathcal{N}}/de) (de/dt)
\sim \dot{\widetilde{\mathcal{N}}}$.
Hence, 

\begin{equation} \label{etransport}
\frac{d\widetilde{\mathcal{N}}}{de} \propto \left(\frac{de}{dt}\right)^{-1}
\propto \frac{e^{3/5}}{(1-e^2)^{3/2}} \,\,\,\, {\rm for} \,\,\,\, e \leq e_0
\, ,
\end{equation}

\ni where we have used (\ref{dedt}).
%
Equation (\ref{etransport})
implies that nearly all barely bound grains have $e \approx e_0$.
%
Therefore
the CPR-evolved eccentricity distribution
closely resembles the initial eccentricity distribution:

\begin{equation}\label{dNde_dNde0}
\frac{d\mathcal{N}}{de} \sim \frac{d\mathcal{N}}{de_0}  \, .
\end{equation}

Now we address the size distribution that determines $d\mathcal{N}/de_0$.
For type A disks, for which
$s_{\rm break} - s_{\rm blow} \gg s_{\rm blow}$, the size distribution
for barely bound grains obeys
 $d\mathcal{N}/ds \propto
s^{-q_{\rm inject}}t_{\rm CPR}(s)$
(see [\ref{theunusual}]).  
For type B disks, $s_{\rm break} - s_{\rm blow} \ll s_{\rm blow}$;
outer-disk grains for which $s_{\rm blow} < s < s_{\rm break}$ are
outnumbered by grains having $s > s_{\rm break}$. The relevant size
distribution in the outer regions of type B disks is therefore
the one appropriate for $s > s_{\rm break}$:
$d\mathcal{N}/ds
\propto s^{-7/2}(1-e_0)^{-3/2}$ (see [\ref{dohn+e1}]).\footnote{This
  statement for type B disks
  is only valid not too far
  from the birth ring. As $r\rightarrow\infty$,
  the only bound grains that are present have
  sizes between $s_{\rm blow}$ and $s_{\rm break}$.  These obey
  $\tau_\perp \propto r^{-5/2}$,
  just as they do for type A disks. Therefore for type B disks,
  the $\tau_\perp \propto r^{-3/2}$ scaling derived in the main text
  eventually gives way to $\tau_\perp \propto r^{-5/2}$. \label{faraway_almost}}
We evaluate these distributions for $s = s_{\rm blow,+}$:

\begin{subequations}
\begin{equation}
\frac{d\mathcal{N}_{\rm blow,+}}{ds} \propto E(e_0) \propto (1-e_0)^{-1/2}
\,\,\,\, {\rm for} \,\, {\rm Type \,\, A \,\, disks} \,,
\end{equation}
\begin{equation}
\frac{d\mathcal{N}_{\rm blow,+}}{ds} \propto (1-e_0)^{-3/2}
\,\,\,\, {\rm for} \,\, {\rm Type \,\, B \,\, disks} \,,
\end{equation}
\end{subequations}

\ni using equations (\ref{tcpr}) and (\ref{Egrowth}). 
Now
$d\mathcal{N}/de \sim d\mathcal{N}/de_0 = (d\mathcal{N}/ds)(ds/de_0)$.  
Since equations (\ref{e0}) imply
that $ds/de_0$ is approximately constant for $e_0 \approx 1$ (i.e., for
$s = s_{\rm blow,+}$), and
since $r \approx r_{\rm apo} \approx 2r_{\rm BR}/(1-e_0)$,
%
%

\begin{subequations}\label{hardest}
\begin{equation}
\frac{d\mathcal{N}_{\rm blow,+}}{de} \propto (1-e_0)^{-1/2} \propto r^{1/2}
\,\,\,\, {\rm for} \,\, {\rm Type \,\, A \,\, disks} \, ,
\end{equation}
\begin{equation}\label{hardestb}
\frac{d\mathcal{N}_{\rm blow,+}}{de} \propto (1-e_0)^{-3/2} \propto r^{3/2}
\,\,\,\, {\rm for} \,\, {\rm Type \,\, B \,\, disks} \,.
\end{equation}
\end{subequations}

%

\ni Combine (\ref{massnumber}), (\ref{chain}), (\ref{dedr}), and
(\ref{hardest}) 
 to obtain

\begin{subequations}\label{sigma2.5}
\begin{equation}
\tau_\perp \propto r^{-5/2}
\,\,\,\, {\rm for} \,\, {\rm Type \,\, A \,\, disks} \,,
\end{equation}
\begin{equation}
\tau_\perp \propto r^{-3/2}
\,\,\,\, {\rm for} \,\, {\rm Type \,\, B \,\, disks} \,.
\end{equation}
\end{subequations}

\ni Note that these scaling relations cannot be obtained by merely
assuming that the disk-integrated size distribution obeys the usual
Dohnanyi relation $d\mathcal{N}/ds \propto s^{-7/2}$.  For type B
disks, for example, the key 
modification arises from the prolonging of the collisional lifetime
due to SWR pressure (i.e., the factor of $(1-e_0)^{-3/2}$ in equation
[\ref{hardestb}]).  Moreover, the scalings are robust against
uncertainties in the size distribution; they do not depend explicitly
on either $q_{\rm ce}$ or $q_{\rm inject}$.
We verify these scalings by numerical experiments in \S\ref{montecarlo}.







\subsection{Unbound Grains $(\beta \ge 1/2)$}
\label{unbound}

The rapid expulsion of unbound ($\beta \ge 1/2$) grains compared to the longer,
CPR-driven orbital decay of barely bound ($1/2 - \beta \ll 1$) grains
suggests that in steady state, unbound grains contribute little to the
surface brightness of the outer disk. On the other hand, unbound
grain velocities are nearly constant with radius---for $\beta \approx 1$,
velocities are approximately equal to their (circular, Keplerian)
birth velocities.
As a result, the optical depth of
unbound grains should roughly obey $\tau_{\perp,{\rm ub}} \propto r^{-1}$
and should eventually
exceed, at some ``cross-over radius,'' the optical depth
of barely bound grains, which scales as $\tau_{\perp,{\rm bb}} \propto
r^{-5/2}$ in type A disks and as $\tau_{\perp,{\rm bb}} \propto r^{-3/2}$ in
type B disks (\S\ref{outerphysics}).
In this section, we estimate the value of the
cross-over radius, $r_{\rm cross}$, and show in the case of AU Mic
that it lies outside the scope of current observations.

Consider a type A disk.
In the birth ring,
the optical depth of barely bound grains exceeds that of unbound grains by

\begin{equation}
\tau_{\perp,{\rm bb}} / \tau_{\perp,{\rm ub}} \sim 
\left. \int_{s_{\rm blow}}^{2s_{\rm blow}} \frac{dN}{ds}
 \, s^2 \, ds \right/ 
\int_{\min(s_V)}^{s_{\rm blow}} \frac{dN}{ds}
\, s^2 \, ds \,.
\label{integrals}
\end{equation}

\ni The smallest unbound grain of interest is the smallest grain
for which $Q_{\rm scat} \sim 1$: $\min (s_V) \approx 0.1 \mum$.
Since $\tau_{\perp,{\rm bb}} \propto r^{-5/2}$ while
$\tau_{\perp,{\rm ub}} \propto r^{-1}$, the cross-over radius is

\begin{equation}
r_{\rm cross} \sim \left( \frac{\tau_{\perp,{\rm
      bb}}}{\tau_{\perp,{\rm ub}}} \right)^{2/3} r_{\rm BR} \,.
\end{equation}

\ni By (\ref{dnds_fe}),

\begin{equation}
\left.\frac{dN}{ds}\right|_{s_{\rm blow} < s < 2 s_{\rm blow}}
\sim D s^{-q_{\rm inject}} 
\left.\Delta t_{\rm reside}\right|_{\rm CPR}
\label{theunusualagain}
\end{equation}

\ni in the birth ring, where $D$ is a constant.
By the same logic that led to (\ref{dnds_fe}),

\begin{equation}
\left.\frac{dN}{ds}\right|_{s < s_{\rm blow}} 
\sim D s^{-q_{\rm inject}} t_{\rm blow}(s) \,
\end{equation}

\ni in the birth ring, where

\begin{equation}
t_{\rm blow}(s) \sim \frac{\sqrt{r_{\rm BR}\Delta r}}{\Omega_{\rm BR}
  r_{\rm BR} \sqrt{\beta}}
\sim 90 \left(
\frac{s}{s_{\rm blow}}
\right)^{1/2} \sqrt{\frac{\Delta r}{r_{\rm BR}}} \yr
\label{tblow}
\end{equation}

\ni is the time for an unbound grain to leave the birth ring,
and $\Omega_{\rm BR} = \sqrt{GM_{\ast}/r_{\rm BR}^3}$.
Numerical evaluation of the integrals in (\ref{integrals})
reveals that

\begin{equation} \label{rcrossanswer}
r_{\rm cross} \sim \left\{ \begin{array}{rl}
900\AU &  \,\,\,\, {\rm for} \,\,\,
\dot{M}_{\ast}/\dot{M}_{\odot} = 10^2 \\ 
200\AU &  \,\,\,\, {\rm for} \,\,\,
\dot{M}_{\ast}/\dot{M}_{\odot} = 10^3 \,.
\end{array} \right.
\end{equation}

\ni If instead type B conditions apply for the AU Mic disk,
then by considerations analogous to those above, 
$r_{\rm cross} \gg 10^3\AU$.
We compare $r_{\rm cross}$ with the maximum radius probed by current
observations---approximately $200 \AU$---to
conclude that
under type A conditions,
unbound grains
contribute at most marginally to the currently observed surface
brightness. Under type B conditions, they contribute negligibly.

\subsection{Unequilibrated Grains ($s_{\rm blow} < s < s_{\rm age}$)}
\label{sage}
Grains on extremely eccentric orbits may have collisional
and CPR lifetimes that exceed the age of the system.  
Can such grains, whose numbers cannot be assessed within our
steady-state model, contribute significantly to the observed surface
brightness of the AU Mic disk?
Such ``unequilibrated grains'' have sizes between $s_{\rm blow}$ and
$s_{\rm age}$, where

\begin{equation}\label{sagedef}
t_{\rm age} = \min[t_{\rm CPR}(s_{\rm age}), t_{\rm col}(s_{\rm age})]
\end{equation}

\ni defines $s_{\rm age}$. We apply (\ref{tcpr}) and
(\ref{tcollessbreak}) to find that

\begin{equation}
\frac{s_{\rm age}}{s_{\rm blow}} \sim \left\{ \begin{array}{rl}
1.006 &  \,\,\,\, {\rm for} \,\,\, \dot{M}_{\ast}/\dot{M}_{\odot} =
1 \\
1.002 &  \,\,\,\, {\rm for} \,\,\, \dot{M}_{\ast}/\dot{M}_{\odot} =
10 \\
1.00003 &  \,\,\,\, {\rm for} \,\,\,
\dot{M}_{\ast}/\dot{M}_{\odot} = 10^2 \\ 
1.000002 &  \,\,\,\, {\rm for} \,\,\,
\dot{M}_{\ast}/\dot{M}_{\odot} = 10^3 \,.
\end{array} \right.
\end{equation}

\ni 
Unequilibrated grains occupy such a narrow range of sizes that they
seem unlikely to contribute much to the total optical depth.  We
ignore the unequilibrated population for the remainder of this paper.


\section{MONTE CARLO MODELLING}
\label{montecarlo}

\subsection{Procedure}

To test several of the analytic results derived in \S\ref{theory},
we model the AU Mic disk by means
of a Monte Carlo simulation. We calculate the geometric optical depth
$\tau_{\perp} (r)$, edge-on surface brightness profile SB$(b)$, and spectral
flux density $F_{\nu}$ and then we compare to observations.
The input parameters are the stellar mass-loss rate
$\dot{M}_\ast$, optical depth in the center of the birth ring
$\tau_{\perp,{\rm BR}} \equiv \tau_\perp(r_{\rm BR})$,
and width of the birth ring $\Delta r$.

We lay down a number $J$ of dust particles
around the central star in a two-dimensional plane. Each particle's
radial coordinate $r$ and azimuth $\psi$ are determined by the
particle's semi-major axis $a$, eccentricity $e$, true anomaly $\phi_{\rm t}$,
and longitude of periastron $\pomega$. Since our model disk is axisymmetric,
$\pomega$ is drawn as a uniform deviate from 0 to $2\pi$.

The birth distributions of the remaining
orbital elements depend on the
distribution of
grain sizes $s$.
Orbital elements subsequently evolve from their birth values
by CPR drag.
The degree of evolution depends on the
age of the grain ($t$) relative to the CPR lifetime ($t_{\rm CPR}$; equation
[\ref{tcpr}])
and collision lifetime ($t_{\rm col}$; equations [\ref{tcolfull}] and
[\ref{tcollessbreak}]).
By definition, $s_{\rm break}$ is the grain
size for which $t_{\rm CPR} = t_{\rm col}$,
and
$s_{\rm age}$ is the smallest grain size for
which $\min(t_{\rm CPR}, t_{\rm col}) = t_{\rm age}$.
Only grains having $s > s_{\rm age}$ can be removed over the age of the system.
For $s > s_{\rm break}$, collisions are more important than CPR
drag in removing grains ($t_{\rm col} < t_{\rm CPR}$), so
we draw the ages $t$ of such particles
as a uniform deviate from 0 to $t_{\rm col}(s)$.
Grains of size
$s < s_{\rm break}$ are removed by CPR drag ($t_{\rm CPR} <
t_{\rm col}$), so
we draw $t$ for these grains as a uniform deviate from 0 to
$t_{\rm CPR}(s)$.

The evolved eccentricity $e$ depends on the initial eccentricity $e_0$
and the age of the particle, implicitly according to

\begin{equation}\label{t(e)}
\frac{t}{t_{\rm CPR}} =
1-\frac{E(e)}{E(e_0)}
\left(\frac{1+e_0}{1+e}\right)^{2}
\left(\frac{e}{e_0}\right)^{8/5} \, ,
\end{equation}

\ni where $E(e)$ is defined by (\ref{WW_Edef}).
The evolved semi-major axis is given by

\begin{equation}
a = a_0 \left( \frac{e}{e_0} \right)^{4/5} 
\left( \frac{1-e_0^2}{1-e^2} \right)
\label{drawa}
\end{equation}

\ni \citep{wyattwhipple}, where the initial semi-major axis
$a_0$ is derived from the initial periastron $r_{{\rm peri},0}$, 
which we draw in the
following way.  In our model, all grains are born in the birth ring,
an annulus of width $\Delta r$ centered at $r_{\rm BR}$. At birth, a grain is
located at the periastron of its osculating orbit---an orbit rendered eccentric
by SWR forces (\S\ref{blowout}).  We draw $r_{{\rm peri},0}$
from a uniform distribution of width $\Delta r$ centered at $r_{\rm BR}$.



As mentioned, 
the birth distributions of eccentricities and semi-major axes
depend on the distribution of grain sizes $s$.  Although particles in
our simulation are
born only in the birth ring, their steady-state population may occupy
all space, so we must draw $s$ from
the global, disk-integrated size distribution $d\mathcal{N}/ds$.  We
apply results from
\S\ref{cascades}, made more precise for our Monte Carlo calculation.
From (\ref{theunusual}), 

\begin{equation}\label{simplednds_small}
\frac{d\mathcal{N}}{ds}
\propto s^{-q_{\rm inject}}t_{\rm
  CPR}(s,r_{\rm peri}=r_{\rm BR})
\,\,\,\, {\rm for} \,\,\,\, s_{\rm age} < s < s_{\rm break} \, ,
\end{equation}

\ni and from (\ref{dohn+e}), 

\begin{equation}\label{fulldnds_big}
\frac{d\mathcal{N}}{ds} \propto s^{-7/2}
\frac{t_{\rm col}}{\left.\Delta t_{\rm reside}\right|_{\rm col}}
\,\,\,\, {\rm for} \,\,\,\, s > s_{\rm break} \, .
\end{equation}

\begin{figure}
\epsscale{1.3}
\plotone{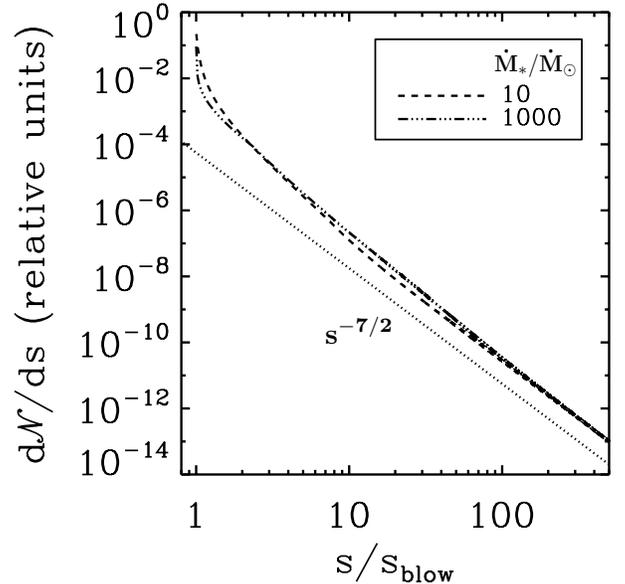}
\caption{Disk-integrated grain size distributions
    $d\mathcal{N}/ds$. 
    Dashed and triple-dot--dashed
    curves correspond respectively to
      $\dot{M}_{\ast}/\dot{M}_{\odot} = \{10, 10^3\}$.
    A dotted line proportional to the Dohnanyi scaling $s^{-q_{\rm
    ce}} = s^{-7/2}$ is
    overplotted for reference.  As $s$ approaches $s_{\rm blow}$, the
    population of grains rises significantly above what one would
    expect from a pure Dohnanyi size spectrum.  The deviations from a
    Dohnanyi spectrum, which differ under type A (e.g., $\dot{M}_\ast =
    10^3\dot{M}_\odot$) and type B (e.g., $\dot{M}_\ast =
    10\dot{M}_\odot$) conditions, are critical for understanding how
    $\tau_\perp$ scales with $r$ in the outer disk; see \S\ref{outerphysics}.
\label{dmathcalN}}
\end{figure}

\ni We approximate $t_{\rm col}$ using 
(\ref{tcolfull}), and
we evaluate $\Delta t_{\rm reside}|_{\rm col}$ numerically.
We take the distribution described by
(\ref{simplednds_small}) and (\ref{fulldnds_big})
to be non-zero only for 
$s > s_{\rm age}$\footnote{For
$\dot{M}_{\ast}/\dot{M}_{\odot} = 1$, $s_{\rm age} > s_{\rm break}$
so only (\ref{fulldnds_big}) is relevant.}
and to be continuous across $s_{\rm break}$; furthermore, we truncate the
distribution at $s_{\rm max} = 500 s_{\rm blow}$ because of
computational limitations.
We fix $q_{\rm inject} = 4$.  The distribution $d\mathcal{N}/ds$ is
plotted in Figure \ref{dmathcalN}.
For each $s$ drawn, $e_0$ and $a_0$
are calculated using equations (\ref{e0}). For $J = 10^9$, one Monte
Carlo simulation takes 8 hr to complete on a 1.33 GHz PowerPC G4
processor.

That the distribution of
mean anomalies $\phi_{\rm m}$ is uniform determines the distribution of
true anomalies $\phi_{\rm t}$ via Kepler's equation \citep{murraydermott}:

\begin{equation} \label{Meqn}
\phi_{\rm m} = \phi_{\rm e} - e\sin \phi_{\rm e} \, ,
\end{equation}

\ni where $\phi_{\rm e}$ is the eccentric anomaly:

\begin{equation} \label{feqn}
\tan \frac{\phi_{\rm t}}{2} = 
\left(\frac{1+e}{1-e}\right)^{1/2} \tan \frac{\phi_{\rm e}}{2}
\, .
\end{equation}

\ni Knowing $a$, $e$ and $\phi_{\rm t}$ for each particle determines its
radial distance from the star:


\begin{equation}
r = \frac{a(1-e^2)}{1+e\cos \phi_{\rm t}} \, .
\end{equation}


\subsection{Products of the Monte Carlo Calculation}
Having laid down $J$ particles of various sizes, 
we output the following:

\begin{enumerate}

\item The geometric vertical optical depth
$\tau_\perp(r)$.  We first calculate this quantity in relative units
by summing
the geometric cross sections of particles in a given annulus,
and dividing the resultant sum by the area of that annulus.  We
then normalize this result
by matching the input $\tau_{\perp,{\rm BR}}$ to the model's relative optical depth at $r_{\rm BR}$.

\item
The surface brightness of the disk observed edge-on at $V$-band
({\it Hubble Space Telescope}'s $F606W$) and
$H$-band wavelengths, as a function of projected stellar separation $b$:

\begin{align}\label{SB}
& {\rm SB}(b, \lambda)  \nonumber \\ 
& = \int\!\!\!
\int \frac{\lambda L_{\lambda,\ast}}{4\pi r^2}
Q_{\rm scat}(\lambda, s) 
\, P(\theta,\lambda, s) \, \pi s^2 \, \frac{dn}{ds}(r,s) \, d\ell \, ds \, ,
\end{align}


\ni where $\ell = \pm\sqrt{r^2-b^2}$ measures distance along our line of sight.
The stellar flux incident on a grain in the wave band of interest is $\lambda
 L_{\lambda,\ast}/4\pi r^2$,
the cross section for scattering is $Q_{\rm scat} \pi s^2$,
the scattering angle between the star, grain, and observer is
$\theta = \tan^{-1}(\ell/b)$, and 
the relative power scattered per steradian is
$P$ (normalized so that its integral over all solid
angle equals unity).
We use Mie theory to calculate $Q_{\rm scat}$
and $P$, adopting the optical constants of pure water ice \citep{warren}.
The volumetric number density of grains $n$ is found by dividing the
geometric vertical optical depth by the height of the disk:

\begin{equation}
\pi s^2 \frac{dn}{ds}(r,s) = \frac{1}{h(r)}\frac{d\tau_\perp}{ds}(r,s) \, .
\end{equation}

\ni The radial height profile $h(r)$ is derived from the
 projected disk height $h(b)$, which
 roughly follows a broken power law that changes slope around
 $b=r_{\rm BR}=43\AU$:

\begin{equation} \label{h(b)}
h(b) = h_{\rm BR} \left\{ \begin{array}{rl}
\left({b}/{r_{\rm BR}}\right)^{\eta_1} & 
\,\,\,\, {\rm for} \,\,\,\, b<r_{\rm BR} \\
\left({b}/{r_{\rm BR}}\right)^{\eta_2} & \,\,\,\,
{\rm for} \,\,\,\, b>r_{\rm BR} \,.
\end{array} \right.
\end{equation}

\ni We set $h_{\rm BR} = 2.75\AU$. \citet{krist} fit separate broken
power laws to the northwest and southeast extensions of the disk
and obtain
$\eta_1 \approx 0$ and $\eta_2 \approx 1$--2; see their
Figure 7.  For simplicity, we take $\eta_1 = 0$ and $\eta_2 = 1$.
These values characterize an inner disk that is empty and seen in
projection (or that has
constant height), and 
an outer disk in
which grains have constant
inclination dispersion.
We adopt a radial profile $h(r)$ identical to $h(b)$ as given in
(\ref{h(b)}) with $b$
replaced by $r$.
We divide the $F606W$ profile by the $H$
profile to obtain a $V-H$ color profile.

\item The spectral flux density
$F_{\nu}$:

\begin{align}
& F_{\nu} = 
\frac{1}{d^2} B_{\nu}(T_\ast)\, \pi R_\ast^2 \nonumber \\
& + \frac{1}{d^2}\int \!\!\!
\int B_{\nu}(T(r,s)) \, Q_{\rm emis}
(\lambda, s) \, \pi s^2 \,
\frac{dN}{ds}(r,s) \, 2\pi r\,dr \, ds \, ,
\end{align}

\ni where $B_{\nu}(T)$ is the Planck function.
We model the emissivity of the dust as a broken power law:

\begin{equation}
Q_{\rm emis} = 
\left\{ \begin{array}{cl}
1 &  \,\,\,\, {\rm for} \,\,\, 2\pi s > \lambda \\
2\pi s/\lambda & \,\,\,\, {\rm otherwise} \, ,   
\end{array} \right.
\end{equation}

\ni in approximate agreement with the model of \citet{chiang} for ice-mantled
silicate grains.  
Since the peak wavelength of emission from AU Mic is about $1\mum$,
all bound grains are large enough to absorb most of the incident
stellar flux.
We solve 

\begin{equation}
\frac{L_\ast}{4\pi r^2} \pi s^2 = 
4\pi \int_0^{\infty} B_{\lambda}(T)Q_{\rm emis}(\lambda,s)\pi s^2
\,d\lambda 
\end{equation}

\ni for the temperature $T$ specific to a given grain size.


\end{enumerate}

\subsection{Results}
\label{MCresults}

By experimenting with various values of $\tau_{\perp,{\rm BR}}$,
$\Delta r$, and
$\dot{M}_{\ast} \in (1,10,10^2,10^3)\dot{M}_{\odot}$,
we find that $\tau_{\perp,{\rm BR}} = 4\times10^{-3}$,
$\Delta r/r_{\rm BR} = 0.1$, and
$\dot{M}_{\ast} \in (1, 10)\dot{M}_{\odot}$ yield theoretical surface
brightness profiles and spectra that agree encouragingly well with
observations.  Our preferred values for $\tau_{\perp,{\rm BR}}$
and $\Delta r$ are likely uniquely determined to within factors of a few;
SB$(b = r_{\rm BR}) \propto \tau_{\perp,{\rm BR}} \sqrt{\Delta r}$
while $F_{\nu} \propto \tau_{\perp,{\rm BR}} \Delta r$.
The close resemblance of the models for which
$\dot{M}_{\ast} \lesssim 10 \dot{M}_{\odot}$
means that we cannot do better than place an upper limit
on $\dot{M}_{\ast}/\dot{M}_{\odot}$ of $\sim$10.
In what follows,
we present results for our preferred input parameters,
in addition to models for which $\dot{M}_\ast \in (10^2,
10^3)\dot{M}_\odot$ to study the effect of varying $\dot{M}_\ast$ alone.


%

\begin{figure}
\epsscale{1.3}
\hspace{-.4in}
\plotone{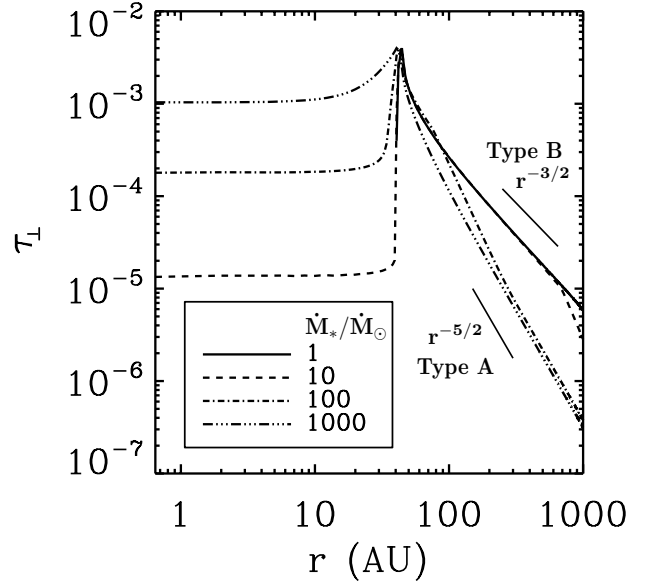}
\caption{Vertical optical depth profiles computed from our Monte Carlo
simulations. 
    Solid, dashed, dot-dashed, and triple-dot--dashed
    curves correspond respectively to
      $\dot{M}_{\ast}/\dot{M}_{\odot} = \{1, 10, 10^2, 10^3\}$.
    Values for $\tau_{\perp,{\rm BR}} = 0.004$ and $\Delta r/r_{\rm
      BR} = 0.1$ are held fixed for all models.
    The two types of disks, CPR-dominated type A disks and
    collision-dominated type B disks, may be distinguished.
The inner disk for
$\dot{M}_\ast = 1\dot{M}_\odot$ is completely empty because $s_{\rm
  break} < s_{\rm age}$.
\label{MCtaufig}}
\end{figure}

Figure \ref{MCtaufig} displays geometric optical depth profiles
$\tau_\perp(r)$.
As expected from our analysis in \S\ref{physics}, disks separate
into two types,
CPR-dominated type A and collision-dominated type B.
As $\dot{M}_\ast$ increases, disk behavior changes from type B to Type
A.  One consequence is that the inner disk becomes increasingly filled
in.  Furthermore, for type A disks, we expect $\tau_\perp \propto
r^{-5/2}$ at $r \gg r_{\rm BR}$; this behavior is indeed evident for
$\dot{M}_\ast/\dot{M}_\odot \in (10^2,10^3)$.
For type B disks, we expect $\tau_\perp \propto r^{-3/2}$ at $r \gg
r_{\rm BR}$; the models for which $\dot{M}_\ast/\dot{M}_\odot \in (1,10)$
exhibit this scaling.

\begin{figure}
  \epsscale{1.25}
\hspace{+2in}
\plotone{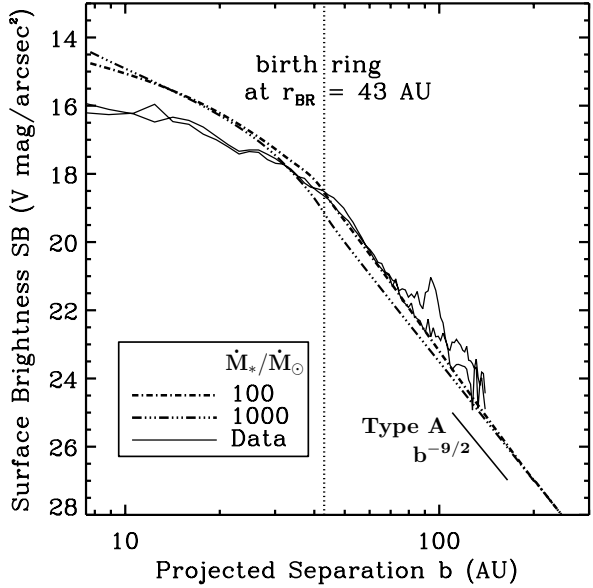}
\caption{Theoretical type A and observed surface brightness profiles.
  Thin lines are data (for northwest and southeast extensions of the disk)
  from
  K05.
    dot-dashed and triple-dot--dashed
    curves correspond respectively to
      $\dot{M}_{\ast}/\dot{M}_{\odot} = \{10^2, 10^3\}$.
  The vertical dotted line corresponds to $r_{\rm BR} = 43\AU$, 
  the radius of the
  birth ring containing dust-producing parent bodies.
The inset scaling of $b^{-9/2}$ is derived from the rule of thumb that
at large $b$, SB $\propto
b^{\gamma-\eta-1}$ for
$h \propto r^\eta$ and $\tau_\perp \propto r^\gamma$.
According to our theory for type A disks, $\eta = 1$ and $\gamma = -5/2$.
\label{MCSBfig_A}}
\end{figure}

\begin{figure}
\epsscale{1.25}
\plotone{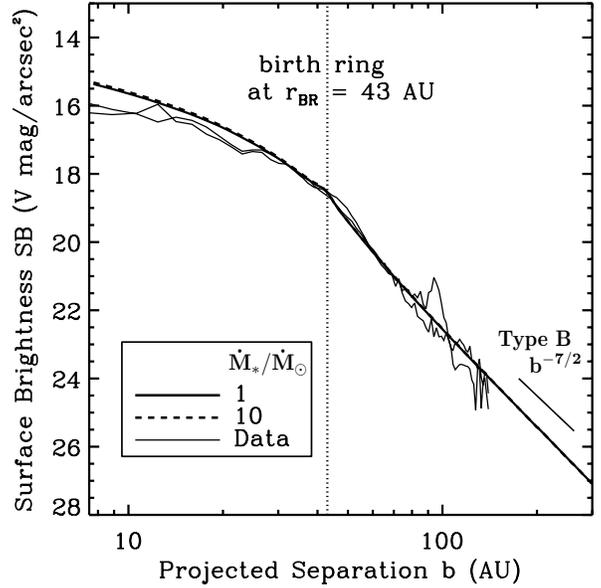}
\caption{Same as Figure \ref{MCSBfig_A} but for type B disk models.
    Thick solid and dashed
    curves correspond respectively to
      $\dot{M}_{\ast}/\dot{M}_{\odot} = \{1, 10\}$.
  The vertical dotted line corresponds to $r_{\rm BR} = 43\AU$, 
  the radius of the
  birth ring containing dust-producing parent bodies.
The inset scaling of $b^{-7/2}$ is derived from the rule of thumb that
at large $b$, SB $\propto
b^{\gamma-\eta-1}$ for
$h \propto r^\eta$ and $\tau_\perp \propto r^\gamma$.
According to our theory for type B disks, $\eta = 1$ and $\gamma =-3/2$.
  Stellar mass-loss rates of 1--10$\dot{M}_\odot$ yield surface
    brightness profiles that agree better with
    the data than those derived from rates of
    $10^2$--$10^3\dot{M}_\odot$; contrast with Figure \ref{MCSBfig_A}.
\label{MCSBfig_B}}
\end{figure}

We compare our theoretical surface brightness
profiles SB$(b)$ with data from K05 in Figures \ref{MCSBfig_A} and
\ref{MCSBfig_B}, for the cases of high $\dot{M}_\ast$ and low
$\dot{M}_\ast$, respectively.
For all disk models, there is a significant contribution to the surface
brightness at $b < r_{\rm BR}$ from starlight that is
forward scattered by grains
located within
the half of the birth ring nearest the observer.
As $\dot{M}_\ast$ increases, the inner disk fills in and the surface
brightness at $b < r_{\rm BR}$ increases.
In comparison, the surface brightness at $b \gg r_{\rm BR}$ decreases with
increasing $\dot{M}_\ast$, reflecting the transition from the
$\tau_\perp \propto r^{-3/2}$ scaling of type B disks to the
$\tau_{\perp} \propto r^{-5/2}$ scaling of type A 
disks. 
Examination of either the inner or outer disk profiles reveals that
models for which $\dot{M}_\ast/\dot{M}_\odot \in (1,10)$ fit the data
better than do models for which $\dot{M}_\ast$ is higher.
Discrepancies between these low-$\dot{M}_\ast$ models and the
observations are less than a factor of 2.  They might arise in
part from our use of a scattering phase function ($P$) appropriate for
idealized spherical grains of pure water ice.

\begin{figure}
\epsscale{1.4}
\hspace{-.7in}
  \plotone{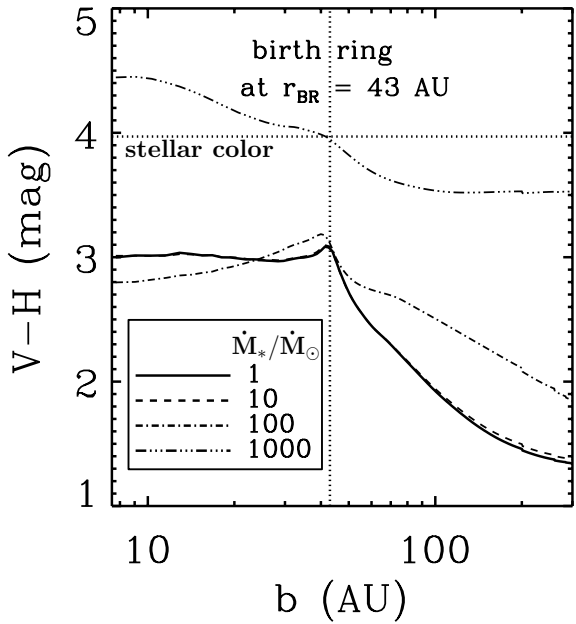}
  \caption{ Color profile ($F606W-H$)
    computed using our theoretical Monte Carlo model.
    Solid, dashed, dot-dashed, and triple-dot--dashed
    curves correspond respectively to
      $\dot{M}_{\ast}/\dot{M}_{\odot} = \{1, 10, 10^2, 10^3\}$.
    The vertical dotted line corresponds to $r_{\rm BR} = 43\AU$, 
    the radius of the
    birth ring containing dust-producing parent bodies;
    the horizontal dotted line is the star's color.
    The outer disk is 
    expected to be bluer in scattered near-infrared light than
    the inner disk.  
    \label{colorplot}}
\end{figure}

In Figure \ref{colorplot}, we plot $V-H$ colors.
For low values of $\dot{M}_\ast$, the inner disk is largely evacuated
and so there is little variation in color
with $b$ for $b < r_{\rm BR}$.  The outer disk becomes progressively bluer
with $b$ as ever smaller (still bound) grains are probed.  All of
these trends are in agreement with observations of disk color
(S.~Metchev 2005, private communication; M.~Fitzgerald 2005, private
communication).

\begin{figure}
\epsscale{1.4}
\hspace{-.7in}
  \plotone{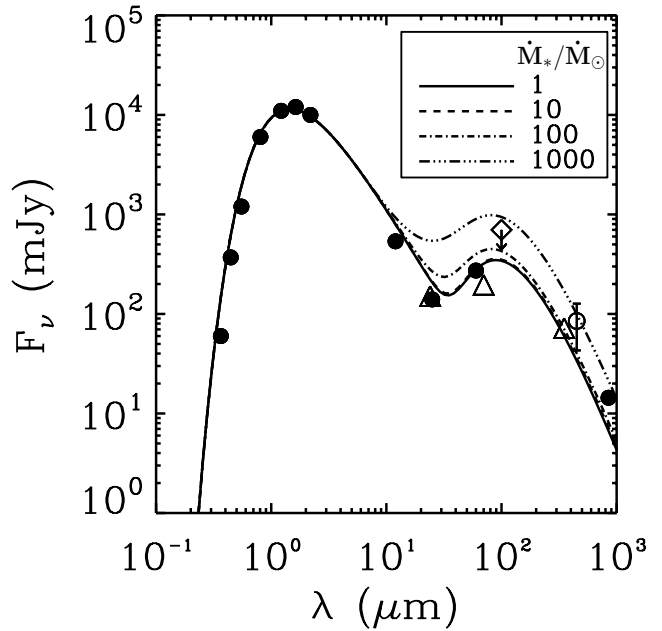}
  \caption{Spectra
    computed using our theoretical Monte Carlo model.
    Solid, dashed, dot-dashed, and triple-dot--dashed
    curves correspond respectively to
      $\dot{M}_{\ast}/\dot{M}_{\odot} = \{1, 10, 10^2, 10^3\}$.
    Data from \citet{liu2} (circles and diamond) and \cite{chen}
    (triangles) are overlaid; the open circle is a possible detection,
    and the diamond represents an upper limit. The discrepancy
    between theory and observation at $\lambda \approx 12 \mum$ is due
    to our use of a blackbody spectrum for the central M dwarf; better
    agreement can be had by employing more realistic stellar atmosphere
    models (e.g., Allard et al.~2001).  
    The discrepancy between the data and the low-$\dot{M}_\ast$ models
    at the longest wavelengths is due at least in part
    to the fact that the maximum grain size in
    our simulations is $500 s_{\rm blow} \ll s_{\rm max}$; were we to
    increase the number of particles $J$ in our simulation, the
    discrepancy would be reduced.
    \label{SEDplot}}
\end{figure}

In Figure \ref{SEDplot},
we plot our theoretical spectra together with
flux measurements from \citet{liu2} and \citet{chen}.
The filled inner disks of high-$\dot{M}_{\ast}$, type A models
produce too much emission at mid-infrared wavelengths to compare
well with observations.
As was our conclusion from studying the surface brightness profile
in reflected starlight, the thermal emission spectra point
to stellar mass-loss rates of $\lesssim 10 \dot{M}_{\odot}$.

\section{SUMMARY AND DIRECTIONS FOR FUTURE WORK}
\label{discussion}

We have constructed a theory to explain the observed optical surface brightness
profile and infrared emission spectrum
of the debris disk encircling AU Mic.  
In our theory, the slope of surface brightness versus projected radius $b$
changes abruptly at $b = 43 \AU$ because a birth ring of planetesimals
exists at stellocentric radius $r = r_{\rm BR} = 43 \AU$.
This ring has a full radial width $\Delta r \sim 0.1 r_{\rm BR}$
and a vertical, geometric optical depth of $\tau_{\perp,{\rm BR}} \sim 0.004$.
The parent bodies in the ring have sizes $s_{\rm max} \sim 10 \cm$
and a total mass of $M_{\rm max} \sim 0.01 M_{\oplus}$.
Collisional attrition of parent bodies
generates micron-sized grains that scatter
starlight at optical wavelengths. The population
of visible grains is maintained in steady state: production by
colliding parent bodies balances removal by grain-grain collisions
and removal by corpuscular and Poynting-Robertson (CPR) drag.
The timescales over which removal of visible grains occurs can be
orders of magnitude shorter than the age of the system ($\sim$12 Myr),
ensuring steady state.

Collisions between parent bodies initiate a collisional
cascade that extends downward in particle size by several orders of magnitude.
Grains having sizes $s < s_{\rm blow} \approx 0.2\mum$
are expelled from the system by stellar wind and radiation (SWR) pressure
and contribute negligibly to the observed optical emission.
Instead, barely bound grains, having sizes just
larger than $s_{\rm blow}$ and which occupy highly eccentric orbits,
make the dominant contribution to the surface
brightness in the outer disk at $r > r_{\rm BR}$.
The number of such grains rises more steeply than would be expected
from a pure Dohnanyi size spectrum as 
$s$ approaches $s_{\rm blow}$ from above, because grains on high-eccentricity
orbits have prolonged lifetimes against CPR drag and collisions.
The structure of the outer disk depends on whether these smallest
of bound grains are removed principally by CPR drag (type A conditions)
or by destructive grain-grain collisions (type B conditions).
As the luminosity and/or mass-loss rate of the central star
increases, disk behavior grades from type B to type A.
As the number of parent bodies in the birth ring increases,
collision rates increase and disk behavior changes from type A to type B.
In the outer reaches of type A disks, the vertical optical depth
scales approximately as $\tau_\perp \propto r^{-5/2}$.
Under type B conditions, $\tau_\perp \propto r^{-3/2}$
(but see footnote \ref{faraway_almost}). We have derived
these scaling relations analytically and have verified
them by Monte Carlo simulations. 

The inner regions at $r < r_{\rm BR}$
are populated by grains that survive long enough before suffering
destructive collisions that their periastron distances diminish
appreciably by CPR drag. In type A disks, a significant fraction of grains
born in the birth ring meet this criterion, so the inner disk is
characterized by
the same vertical optical depth that characterizes the birth ring.
By contrast, under type B conditions, the inner disk is practically empty.

In the case of AU Mic, type B conditions prevail.
By fitting simultaneously both the surface brightness profile and the
thermal emission spectrum, we not only uniquely determine the vertical optical
depth and radial width of the birth ring (see the values cited above), but also
constrain the stellar mass-loss rate $\dot{M}_\ast$ to be
$\lesssim 10\dot{M}_\odot$.
According to our theory, the inner disk of AU Mic at $r < r_{\rm BR}$
is empty. The observed surface brightness
at $b < r_{\rm BR}$ is not zero because we are observing the disk edge-on.
The primary contribution to the surface brightness at $b < r_{\rm BR}$
arises from starlight that is forward scattered by grains in the birth ring.

Our theory states that the observed structure of the AU Mic disk
reflects processes that are balanced in steady state.
Equilibrium is likely since the timescales over which collisions and
CPR drag operate, even in the rarefied outer disk,
are shorter than the age of the system.
The outer disk does not comprise ``primordial'' grains left
behind from a now-evaporated gaseous disk,
as has been speculated previously.
Nor is the manifestation of the debris disk phenomenon in AU Mic
the outcome of a recent cataclysm that has not yet equilibrated.
That our required parent body mass is modest
(equation [\ref{parentmass}])
supports our contention that the AU Mic disk is in steady state.

As our paper was being completed, we became aware of an independent
study of AU Mic by \citet{auger06}. These authors find by
inverting the observed surface brightness profile that the underlying
vertical optical depth of the AU Mic disk peaks near 35 AU.
It is heartening that
their conclusion is consonant with one of ours, derived as they are
using complementary approaches:
detailed data-fitting procedures versus
physical reasoning to understand dust dynamics under general circumstances.

The ring of parent bodies at $r_{\rm BR} = 43 \AU$ that we envision
encircling AU Mic presents a youthful analogue to
the Solar System's Kuiper belt
(see the Protostars and Planets V review by Chiang et al.~2006).
The spatial dimensions of these systems are remarkably similar: The
Classical Kuiper belt, containing
those planetesimals thought to have formed {\it in situ},
extends in heliocentric distance from $\sim$40 AU to $\sim$47 AU
(e.g., Trujillo \& Brown 2001).

We conclude by pointing out directions for future work on AU Mic and other
debris disks.
\begin{enumerate}

\item {\it Disk thickness.}---By assuming that we are viewing the AU Mic disk
perfectly edge-on, we estimated a full disk height 
of $h_{\rm BR} \approx 2.75 \AU$. The corresponding opening angle
is $h_{\rm BR}/r_{\rm BR} \approx 4^{\circ}$. While our model of a type B
disk succeeds in reproducing the observed
scaling behavior of disk height ($h \propto b^0$ in the inner disk
and $h \propto b^1$ in the outer disk), we have not explained what sets
the normalization. Dissipative grain-grain collisions would be expected
to damp the inclination dispersion and to reduce $h_{\rm BR}$
to values orders of magnitude smaller than our inferred value.

\item {\it Application to other systems.}---The 
  debris disk surrounding the A-star
  $\beta$ Pictoris closely resembles the AU Mic disk
  \citep{liu}. The surface brightness profile abruptly 
  changes slope at $b \approx 100 \AU$,
  from SB $\propto b^{-2.4}$ to SB $\propto b^{-4.0}$ (Kalas \& Jewitt 1995).
  Moreover, the vertical scale height $h$
  scales with $b$ the same way that it does in the AU Mic
  disk.\footnote{That the color of $\beta$ Pic's outer disk is red rather
    than blue (by contrast to the case of AU Mic) could
    be a consequence of the particular grain size required for blow-out
    in the $\beta$ Pic system, since
  for certain grain sizes and compositions,
  $Q_{\rm scat}(\lambda, s)$ can actually increase with increasing wavelength.
  See Bohren \& Huffman (1983) for a discussion of this phenomenon of
  ``blueing.''}
  Recently, another analogue to the AU Mic disk has been discovered: F-star
  HD 139664 hosts a debris disk whose surface brightness profile 
  exhibits a sharp break at $\sim$$90\AU$ \citep{kalas2006}.
  The theory we have laid out for AU Mic might find ready application
  to these other systems.
  
\item {\it Uniqueness of AU Mic among M dwarfs.}---The pioneering 
  Keck survey
  conducted by PJL05 at $\lambda = 11.7 \mum$ reveals that AU Mic 
  is distinguished among their sample of
  nine M dwarfs having ages of 10--500 Myr
  in emitting an infrared excess. Why? Do the other M dwarfs not possess disks?
  As M dwarfs constitute the most numerous
  stars in the universe, understanding why AU Mic might be exceptional
  will help to determine the prevalence of planetary systems.
  Many of the M dwarfs surveyed by PJL05 may simply be much older than
  AU Mic; their parent body populations may have suffered near
  complete comminution.
  
\item {\it Ubiquity of rings.}---That
  parent bodies are confined to a ring centered at $43\AU$ in the
  AU Mic system calls for explanation. Ring morphologies are so
  common---witness the examples of HR 4796A (Schneider et al.~1999),
  $\epsilon$ Eridani (Greaves et al.~1998), Fomalhaut
  (Kalas et al.~2005), and even the Kuiper Belt---that the
  ``debris disk phenomenon'' might
  well be more precisely termed the ``debris ring phenomenon.''
  While regions interior to rings might have been purged
  of material by planets, the physical processes that determine
  the outer edges of rings remain unclear.
  Ideas proposed by Takeuchi \& Artymowicz (2001)
  and Klahr \& Lin (2005) for how interactions between solids and gas
  can concentrate planetesimals into rings might be relevant.
  That planetary systems have sharp outer edges suggests that
  planetesimal formation is not a continuous function of disk properties;
  rather, the formation of planets may require
  disk properties to meet threshold conditions (e.g., Youdin 2004).


\end{enumerate}

\acknowledgements
This work was made possible by grants from the National Science Foundation
and the Alfred P.~Sloan Foundation. We are grateful to Peter Plavchan for
extensive and helpful discussions, and to John Krist for supplying us
with {\it HST} surface brightness data.  We acknowledge encouraging exchanges
with Pawel Artymowicz, Doug Baker, Josh Eisner, Mike Fitzgerald, 
James Graham, Lynne Hillenbrand, Mike Jura, Paul Kalas, Yoram
Lithwick, Holly Maness, Stan Metchev, Re'em Sari, and Yanqin Wu.
A portion of this work was completed in Awaji Island, Japan, in the
cheerful company of the participants of the 2005 Kobe International
Planetary School.

\newpage
\clearpage
\begin{deluxetable}{ccccccc} 
\tabletypesize{\scriptsize}
\tablewidth{0pt}
\tablecaption{Stellar Wind and Radiation Parameters
in the AU Mic System\tablenotemark{a}
\label{windtable}}
\tablehead{
\colhead{$\dot{M_{\ast}}/\dot{M}_{\odot}$} &
\colhead{$P_{\rm SWR}$} & 
\colhead{$P_{\rm CPR}$} &
\colhead{$s_{\rm blow} (\mum)$\tablenotemark{b}} &
\colhead{$t_{\rm CPR} (s = 1.1 s_{\rm blow}) (\yr)$\tablenotemark{c}} &
\colhead{$t_{\rm CPR} (s = 1.5 s_{\rm blow}) (\yr)$\tablenotemark{c}} &
\colhead{$t_{\rm CPR} (s = 15 s_{\rm blow}) (\yr)$\tablenotemark{c}}
}
\startdata
$1$ & 2.0 & $5.0$ & $0.23$ & $9.5 \times 10^6$ & $4.9
\times 10^6$ & $1.9 \times 10^7$ \\
$10$ & 2.0 & $32$ & $0.23$ & $1.5 \times 10^6$ & $7.8
\times 10^5$ & $3.1 \times 10^6$ \\
$10^2$ & 2.4 & $3.0 \times 10^2$ & $0.28$ & $1.9 \times 10^5$ & $9.9
\times 10^4$ & $3.9 \times 10^5$ \\ 
$10^3$ & 6.4 & $3.0 \times 10^3$ & $0.74$ & $5.1 \times 10^4$ & $2.6
\times 10^4$ & $1.0 \times 10^5$ \\
\enddata
\tablenotetext{a}{Assumes $Q_{\rm rad} = 2$, $Q_{\rm wind} = 1$, $\rho = 2 \gm \cm^{-3}$, $v_{\rm wind} = 450 \km \s^{-1}$, $L_{\ast} = 0.1 L_{\odot}$, $M_{\ast} = 0.5 M_{\odot}$, $\dot{M}_{\odot} = 2 \times 10^{-14} M_{\odot} \yr^{-1}$,
and $r_{\rm peri} = r_{\rm BR} = 43 \AU$.}
\tablenotetext{b}{The blow-out radius is such that $\beta = 1/2$.}
\tablenotetext{c}{Calculated for $e = e_0$.
For $s/s_{\rm blow} = \{ 1.1,1.5,15 \}$,
$e_0 = \{ 0.83, 0.50, 0.034 \}$ by equations (\ref{e0})
and $E(e_0) = \{  7.2, 2.7, 1.1 \}$ by equation (\ref{WW_Edef}).}
\end{deluxetable}

\clearpage


\end{document}